%% file: ms.tex
\documentclass[preprint,11pt]{aastex}

\shorttitle{ONC Kinematics}
\shortauthors{Tobin et al.}

\newcommand{\msun}{\mbox{$M_{\sun}$}}

\begin{document}

\title{Kinematics of the Orion Nebula Cluster: Velocity Substructure and Spectroscopic Binaries \footnotemark}
\author{John J. Tobin\altaffilmark{2}, Lee Hartmann\altaffilmark{2}, Gabor Furesz\altaffilmark{3}, Mario Mateo\altaffilmark{2}, 
   \& S. Tom Megeath\altaffilmark{4}}

\begin{abstract}
We present a kinematic study of the Orion Nebula Cluster 
based upon radial velocities measured by multi-fiber echelle
spectroscopy at the 6.5 meter MMT and Magellan telescopes. 
Velocities are reported for 1613 stars, with multi-epoch
data for 727 objects as part of our continuing effort to detect and analyze
spectroscopic binaries.  We confirm and extend the results of Furesz
et al. showing that the ONC is not relaxed, consistent with
its youth, and that the stars generally
follow the position-velocity structure of the moderate density gas in the region, 
traced by $^{13}$CO.
The additional radial velocities we have measured enable us to probe
some discrepancies between stellar and gaseous structure which can be 
attributed to binary motion and the inclusion of non-members in our kinematic
sample.
Our multi-epoch data allow us to identify 89 spectroscopic binaries; more
will be found as we continue monitoring.  Our results reinforce the idea that
the ONC is a cluster in formation, and thus provides a valuable testing ground for theory.
In particular, our observations are not consistent with the quasi-equilibrium
or slow contraction models of cluster formation, but are consistent with cold
collapse models.

\end{abstract}

\keywords{stars: formation}
\footnotetext[1]{This paper includes data gathered with the 6.5 meter Magellan Telescopes located at Las Campanas Observatory, Chile; 
Observations reported here were obtained at the MMT Observatory, a joint facility of the Smithsonian Institution and the University of Arizona.}
\altaffiltext{2}{Department of Astronomy, University of Michigan, Ann Arbor, MI 48109; jjtobin@umich.edu}
\altaffiltext{3}{Center for Astrophysics, 60 Garden Street, Cambridge, MA 02138}
\altaffiltext{4}{Department Astronomy, University of Toledo, 2801 West Bancroft Street, Toledo, OH 43606}

\section{Introduction}
Observational evidence accumulated over the last decade has shown that most
 stars form in clusters \citep{carpenter2000, ladalada2003, allen2007}. Therefore,
understanding cluster formation is an important aspect of star formation theory.
Cluster origin theories range from highly dynamic models \citep{bonnell2003}
to quasi-equilibrium and/or slow contraction scenarios
\citep{tkm2006}. 
Observations of very young clusters, before they have
dynamically relaxed, can provide indications of initial conditions of
formation and thus constraints on theories of cluster formation.

The Orion Nebula Cluster (ONC) is the nearest young cluster of substantial size,
with approximately 2000 $\msun$ of stars within 2 pc of the Trapezium
\citep[][HH98]{hh1998}.  Although HH98 were able to fit a King cluster
model to the azimuthally-averaged spatial distribution of the stars, they suggested
that the ONC may not be fully relaxed. Also, they pointed
out that the cluster becomes increasingly elongated on larger scales, in the direction
of the filamentary molecular gas in the region.  \citet{scm2005} emphasized the elongated 
nature of the stellar distribution on large scales and pointed
out that the King model implied an unrealistically large tidal radius.

The numerical experiments of \citet{scm2005} led to preferred models for the ONC in which 
it might be expanding quasi-statically, though they could not rule out the possibility
that it is in a stage of collapse.  The investigation by
\citet{kroupa2000} similarly could not distinguish between expansion, equilibrium, and collapse
models for the ONC. Also, neither \citet{kroupa2000} or \citet{scm2005} considered the mass
of the molecular cloud in their studies, which is at least equal to the mass of stars.
  \citet{fiegelson2005}
argued that the ONC must be in a state of violent relaxation, noting  significant
 asymmetry in the spatial distribution of young stars in the inner regions
of the cluster.  In contrast, \citet{tkm2006} argued that the ONC and 
other clusters cannot form by global collapse, but instead 
form over several dynamical timescales.  In a similar vein,
\citet{huff2007} argued in favor 
of a model of the ONC where it has been slowly contracting for roughly 10 Myr.

In the first paper of this series, \citet{furesz2008}, we found spatially coherent
kinematic structure in the stellar distribution of the ONC, a structure which
closely matches that of the molecular gas in the region.  This correlated
structure shows that the cluster is not fully relaxed.  Furthermore, we argued
that the cluster is less than a crossing time old; otherwise the gas would have
shocked and dissipated much of its structural relation to the stars.
This inference of unrelaxed structure is consistent with the relative youth of most of the
cluster stars \citep{hill1997}.  In addition, the kinematic observations were
consistent with the cold collapse model for the overall structure of the Orion
A cloud developed by \citet{hb2007}.

In this paper we present additional radial velocity observations of the ONC cluster
stars.  The combined data set allows us to make a more refined comparison between
gaseous and stellar kinematics as a function of position, and to begin characterizing
the spectroscopic binary population of the cluster.  While the overall correlation
between stellar and gas motions remains, we find some departures in detail which
in some cases appear to be the result of blowout of molecular gas by the massive
stars in the region.  Finally, our initial
detections of spectroscopic binaries reveals a number of objects with infrared
excess emission from a disk, a somewhat surprising result as binary companions are thought
to help disrupt circumstellar disks.

\section{Observations and Data Reduction}

We have obtained multi-fiber echelle spectra using Hectochelle \citep{szent1998} on the MMT
and MIKE Fibers \citep{walker2007} on the Magellan Clay telescope. 
Hectochelle uses robots to position all 240 fibers on a 1$^{\circ}$ field of view (FOV), 
while MIKE Fibers uses 256 fibers manually plugged into a pre-drilled plate
on a 25$^{\prime}$ FOV. There are limitations of fiber spacing in a given configuration,
no targets may be closer than 30$^{\prime\prime}$ for Hectochelle and 14$^{\prime\prime}$ for MIKE fibers.
The resolution of Hectochelle is R $\sim$ 35000, approximately twice that of MIKE Fibers. 

We took spectra around the magnesium triplet near 5200\AA. 
The Hectochelle spectra cover the wavelength range of $\sim$5150 - 5300\AA, 
while the MIKE Fibers spectra cover $\sim$5150 - 5210\AA. 
Unlike Hectochelle, MIKE fibers has two independent spectrograph channels; 
at 5200\AA we are able to use both channels with 128 fibers going to 
each channel.  The two channels are essentially separate 
spectrographs with different gratings, optics, and CCDs in the 
same enclosure. 
With Hectochelle, all 240 fibers go through the same light path, 
but the detector uses 2 CCDs with 120 apertures illuminating each CCD.

To explore the star and gas kinematic relationship, \citet{furesz2008}
selected targets in the ONC region drawn initially from the 2MASS catalog.
The initial selection criteria were that the stars have 11.5 $<$ J $<$ 13.5 and 
0.2 $<$ (H - K) $<$ 0.5 in order to avoid heavily extincted stars. These criteria would 
exclude young stars with infrared excess emission from a disk, thus young stars with IR excess
emission were selected from the \textit{Spitzer Space Telescope} survey of the ONC (Megeath et al. in prep.).
In all, 1264 stars were selected from 2MASS and 349 were selected from the the \textit{Spitzer} observations.
It is worth pointing out that the stars selected from 2MASS do overlap with those observed in optical studies 
of the ONC \citep[e.g.][]{hill1997, rebull2001} and we have correlated our targets with the optical catalogs.

This study continues the monitoring of the targets observed by \citet{furesz2008} with the addition
of $\sim$200 new targets from the MIKE Fibers observations.
The targets are plotted  in the left panel of Figure \ref{pv} with the $^{13}$CO map from \citet{bally1987}. 
The stellar density on the main filament is very high; thus due
to limitations of fiber spacing we cannot observe all stars in the densest regions with a single pointing of the MMT.
Additional targets were selected as possible members from their NIR colors to fill the remaining 
fibers in the Hectochelle FOV. The MIKE Fibers fields were positioned in areas of high stellar density to observe the most 
targets with a single configuration.

\subsection{MIKE Fibers Observations}

We observed the ONC with MIKE Fibers 
in January 2007 and March/April 2008 (Table 1). 
In the first run, only 3 of our 4 fields were observed; 
during the second run we obtained data for all 4 fields
and observed one field twice.  Field A is located south 
of the Orion Nebula, Fields B and C partially overlap and are located on the central ONC, and Field D
is located north of the Orion Nebula, on NGC 1977
(Table 1). Due to strong nebular contamination (Paschen continuum emission), many of the spectra in 
Field B were unusable.  Most targets in Fields B, C, and D overlap with targets observed by Hectochelle in \citet{furesz2008},
Field A observed many new targets not previously observed with Hectochelle due to its location 30$^{\prime}$ south of the Trapezium.

The data were reduced using the Image Reduction and Analysis Facility (IRAF)\footnote{IRAF is distributed by the National
Optical Astronomy Observatories,
which are operated by the Association of Universities for Research
in Astronomy, Inc., under cooperative agreement with the National
Science Foundation.} task `ccdproc', 
subtracting the overscan, trimming, subtracting the bias and combined our individual
exposures with `imcombine' set for cosmic ray rejection. 
The spectra were extracted using `twodspec', by first tracing 
the apertures in a quartz flat field frame to create a template of aperture traces to extract the science spectra. 
Due to weak Thorium-Argon (Th-Ar) lamps (in the early run), 
the dispersion correction was applied using the the twilight solar spectrum for 
all fields except C-2 and C-3. The dispersion solution was then corrected for drift 
during the night by cross-correlating the faint Th-Ar spectra taken with the twilight
spectra and the science frames using 'fxcor'. 
Fields C-2 and C-3 were calibrated using newly installed Th-Ar lamps which produced much better wavelength solutions.

\subsection{Hectochelle Observations}

In late October 2007, we observed the ONC with Hectochelle. Two epochs of 
Fields 1 and 2 were observed, and one epoch of field 3 (Table 2). 
Field 1 was centered on NGC 1977 and Fields 2 and 3 were centered 
on the central ONC. All the targets in these fields
were previously observed by \citet{furesz2008}. Due to the crowded field,
 all of the 240 fibers could not be placed on targets; thus, a number of fibers were 
allocated for sky observations. These apertures were used
for subtraction of scattered moonlight and
nebular background. In order to subtract the sky observations, 
it was necessary to normalize the fiber 
throughputs using the twilight solar spectrum, and then subtract the 
average spectrum taken from all the sky fibers.  The data were reduced
using an automated IRAF pipeline developed by G. Furesz which utilizes 
the standard spectral reduction procedures, 
similar to our MIKE fibers procedure. A more detailed description
 of Hectochelle data reduction can be found in \citet{aurora2006}. 

\subsection{Radial velocity measurements}

To measure the radial velocities of our 
observed targets, we used the IRAF package 'rvsao' \citep{mink1998}. This package
determines the radial velocity of an object by cross-correlating the observed 
spectrum with a template of known velocity. 
The quality of the subsequent cross-correlation or signal-to-noise (S/N)
is given by R, defined as
\begin{equation}
R = \frac{h}{\sqrt{2}\sigma_a}
\end{equation}
where $h$ is the height of the peak in the correlation function, 
and $\sigma_a$ is the estimated error from the rms of the antisymmetric
portion of the correlation function. The error in the velocity measurement is then dependent
on the instrument used and is of the form
\begin{equation}
\sigma_v \sim \frac{C}{1+R}
\end{equation}
where $C$ is 20 km s$^{-1}$ for MIKE Fibers and 14 km s$^{-1}$ for Hectochelle based on adding random noise to spectra
as in \citet{hartmann1986}.

For the stellar templates, we used libraries of synthetic stellar spectra 
rather than observed templates.  The use of
synthetic templates enables us to explore a 
wider range of stellar parameters than a few observed templates. For the
MIKE data, we used the spectral library of \citet{munari2005}; for the Hectochelle,
 we used the library by \citet{coelho2005}.  
We used different libraries because the 
\citet{munari2005} templates had a resolution of R $\sim$ 20,000 and thus were
a better match for the MIKE data; 
the \citet{coelho2005} library were R $\sim$ 100,000, and were more appropriate for 
the higher resolution of Hectochelle. We tested the MIKE data with both the 
\citet{coelho2005} and \citet{munari2005} templates and found that 
no systematic velocity shift was introduced by using a particular set of templates. 
In all cases we
used a set of templates with surface gravity log(g) = 3.5,
effective temperature (T$_{eff}$) ranging from 3500 - 7000K in steps of 250K, and Solar metallicity. We
found that there was little need to explore a wider parameter space 
with the templates as the most important factor determining the 
quality of a correlation was the T$_{eff}$ of the template. 

The velocity and template with the highest R 
is selected and matched to the appropriate target coordinates and stored in a 
Starbase database \citep{roll1996}. 
For each target, the coordinates are combined into a truncated 2MASS ID number 
(2MASSID), throwing out fractional seconds in right ascension and declination. 
We then used the 2MASSID to match targets
with previous observations and photometric catalogs. 

For the purposes of comparing
the velocity structure of the stars and gas, we have converted our heliocentric
radial velocities of each target to the kinematical local standard of rest (LSR) velocities \citep{kerr1986}.
We only use the LSR velocities in plotting, but give heliocentric radial velocities in our tables.

\subsection{Zeropoint Shifts}

To compare with the previous Hectochelle studies, it was necessary to correct for zeropoint velocity shifts
due to differences in calibration schemes, observations at differing wavelengths
(the earlier studies were based on spectra near H$\alpha$), as well as temperature
variations within the spectrograph \citep{aurora2006,furesz2006,furesz2008}. 
There is also a fiber-to-fiber velocity offset within Hectochelle because the calibration
lamps do not illuminate the fibers in precisely the same way as astronomical objects.
The fiber-to-fiber offsets have been well-characterized and applied to our data; however, 
these corrections are small compared with overall zeropoint shifts at differing epochs.

We adopted the results of \citet{furesz2008} as baseline velocities to shift our observations to match.
Though, the observations in \citet{furesz2008} were observed in 2 epochs, they ensured that there was enough
overlap between epochs to shift all targets to a common zeropoint established by their observations in 2005.
Their 2005 epoch was chosen as the zeropoint because those observations were taken with the then new calibration system
which is still in use. For Hectochelle, we applied a constant shift for each field given by 
the mean of a gaussian fit to the distribution of radial velocity differences for each target in a field.
The zeropoint shifts for the Hectochelle fields are of order 1 km s$^{-1}$, are given in Tables 2. 
However, many stars in the MIKE Fibers fields were not observed with Hectochelle and few of the 
overlapping stars had velocity measurements with R $>$ 6. Instead, we shifted the mean velocity, 
determined by fitting a gaussian, of the stars in a MIKE Fibers field to them mean velocity 
of stars observed with Hectochelle and within the MIKE Fibers field but not necessarily observed 
with MIKE Fibers. More simply, we used the cluster velocity to correct the zeropoint rather than individual
velocities. The zeropoint variation of the MIKE Fibers (Table 1) fields tend to be larger than those of Hectochelle, 
likely due to instrumental differences. Though, the shifts for fields A-1 and A-2 are curiously larger than the rest.
The error quoted for each zeropoint shift
is the uncertainty in the mean of a gaussian fit to the velocity distribution in each field.

\subsection{Identifying Spectroscopic Binaries}
In addition to precise kinematics, we are able to use the multi-epoch data to
identify spectroscopic binaries (SBs) and reduce their 
effect on the dispersion of stars relative to gas.
To identify the SBs, we follow a standard $\chi^2$ 
method \citep[e.g.][]{hartmann1986, maxted2008} to determine radial velocity variability. 
For each star, we start by calculating the error-weighted average radial velocity and
and standard deviation of the average from the entire set of velocities measured
by MIKE Fibers or Hectochelle.
We require that the R of a radial velocity measurement be $>$ 6.0 to be used in the average
 and $\chi^2$. We take this somewhat stringent limit of S/N
 because we do not want to erroneously identify binary stars due to 
an inaccurate velocity measurement. We also employ this cutoff in R for our kinematic study
for the same reasons and it ensures that our velocity errors will be less than the overall cluster
velocity dispersion.
Each individual measurement error has two components, one is the random error from the 
cross-correlation (Eq. 2) and the second is error associated with the zeropoint velocity shift listed
in Tables 1 \& 2. There maybe some additional sources of systematic error which are dependent
on the light path through the spectrograph and because this changes with temperature, it is difficult
to characterize.

Once we had the average velocity and measurement errors we calculated the reduced $\chi^2$
\begin{equation}
\chi^2_r = \frac{1}{N - 1}\sum_{i}^{N}\frac{(RV_i - \overline{RV})^2}{\sigma_i^2}.
\end{equation}
We then calculated the $\chi^2$ probability and counted any stars with P $<$ 0.0001 ($\sim$3.9$\sigma$) as 
SBs and removed them from the kinematic sample. It is unlikely that we have falsely identified a binary 
as our total sample size is 1613; however, this assumes that we have accurately determined the associated errors.
For those that are not variable, the average radial velocity is then used in our kinematic
study. The multi-epoch observations not only enable filtering of binaries but they also yield a 
more accurate final radial velocity than single measurements alone. Additionally,
we detect SBs from the double-peaked appearance of their correlation function. However,
we only flag the obvious double-lined binaries (SB2s) which have clear double-peaked correlation functions
because quantitative detection by RV variability is more reliable. 
We have removed certain and potential SB2s from the kinematic sample, but we only 
include the SBs detected through RV variation and certain SB2s in our analysis of binary properties.

\section{Results}
We present new velocity measurements for 1124 stars in the ONC with our Hectochelle
and MIKE Fibers data combined.
In Table 3, we give a summary table of all targets used in this study. We have combined
our velocity measurements with those of \citet{furesz2008} and give an average velocity
for each target. This table also includes targets that we have not re-observed but are still
included in the kinematic study. In addition, we include the available optical photometry and
have flagged the detected binaries. The rest of the tables give the 
individual velocity measurements per field for easier identification of observation epoch. 
Hectochelle fields 1, 2, and 3 are detailed in Tables 4, 5, and 6 respectively and  
MIKE Fibers fields A, B, C, and D are detailed in Tables 7, 8, 9, and 10 respectively. 
We include velocities with R $>$ 3.5 in the tables, but in our kinematic study we only 
use velocities determined with R $>$ 6.0 to ensure the robustness of our results.

\subsection{Spectroscopic Binaries}

With the addition of these recent observations, we have a built a multi-epoch dataset
that can be used to perform an initial characterization of the close binary population in the ONC.
The binary population for main sequence solar-type stars has been well determined
by \citet{dqmay1991} and because most stars  form in clusters \citep{ladalada2003}, 
we would expect the ONC and the main-sequence to have similar binary fractions. 
Thus far, studies have shown that the fraction of wide binaries ($>$40 AU separation) 
in the ONC is comparable to the main sequence \citep{duchene1999} or perhaps slightly deficient \citep{reipurth2007}.
However,  \citep{reipurth2007} finds that the ONC is certainly deficient in wide binaries
compared to regions of isolated star formation (e.g. Taurus, Upper Sco).
In contrast to the studies of wide binaries, little work has been done in constraining the binary fraction
of the ONC at separations less than 60 AU using radial velocities.

Thus far, we have detected 89 certain spectroscopic binaries from radial velocity variability
and double peaks in the correlation function. In addition, we list 48 other stars as possible
spectroscopic binaries; they are not included them in our binary analysis but have been removed
from the kinematic sample. Of the certain spectroscopic binaries, 74 are identified from radial velocity
variability out of 727 stars with multiple observation epochs and 15 are SB2s identified
from the double-peaked appearance of their correlation function.
Also, 18 of the 74 SBs identified from velocity variability are
found to be SB2s. The binaries detected by radial velocity variations are listed in Table 11 and the double-peaked correlation 
binaries are listed in Table 12.
 \citet{furesz2008} identified 53 stars as spectroscopic binaries from double-peaked correlation 
functions (27) and velocity variations (26). We do not carry over
the possible binary detections of \citet{furesz2008} into our study as our data are better suited for binary identification
with more epochs and a better region of the optical spectrum. However, we do confirm 4 from velocity variability
but none selected by the presence of a double-peaked correlation function. We do not confirm most binaries from velocity variations
 in because \citet{furesz2008} only required R $>$ 4.0 for a reliable velocity; we require R $>$ 6.0. In all, we have multi-epoch
velocities for 727 targets: 333 with 2 epochs, 265 with 3, 93 with 4, 22 with 5, 9 with 6 and 5 with 7. 

In order to constrain a binary orbit spectroscopically, we must determine 6 parameters: period, 
angle of ascending node, eccentricity, time of periastron passage, velocity barycenter, and velocity amplitude of primary and
secondary. Inclination and angle of the line of nodes cannot be
determined by spectroscopic observations alone, and the masses and semi-major axes are
related to the velocity amplitudes and eccentricities as outlined in \citet{batten1973}. Furthermore, in most cases the detected
binaries are single-lined, meaning we can only determine the orbital elements for the primary and an independent constraint
is needed for the masses. Given the number of parameters and epochs we have available, we cannot
reasonably constrain the orbital parameters of any binaries in our sample. In fact, cannot claim to have
detected even 1 period for most stars. We are also limited by the fact that most binaries with periods
longer than 10 days have eccentric orbits and they spend most of their time at low velocities,
decreasing detectability.

Given these difficulties, we cannot estimate our completeness with a any degree of certainty; however, we can compare our
observed binary fraction to the main-sequence. We have determined the total binary fraction in our sample to be
11.5\% and we estimate that we can detect a binary system with a period out to 4000 days.
\citet{dqmay1991} find that $\sim$17.5\% of stars have a binary companion with a period less
than 4000 days. Thus, our study is \textit{at most} $\sim$65\% complete, but this estimate quite optimistic.
At present, this incompleteness is not problem as we are not attempting to constrain the binary frequency of the ONC
as of yet. Our goal is to remove the most obvious binaries from kinematic dataset to reduce the dispersion they introduce to our
overall measurements of the cluster velocity structure.

Despite the limitations of our binary results, we correlated our target sample with the most recent visual binary study of
 the ONC by \citet{reipurth2007}. Of the 72 binary stars detected by \citet{reipurth2007}, 44 were
observed in our radial velocity study. However, we only identify the stars 
0535179-051532 (JW 560) and 0535254-053403 (JW 783) as  
binaries in our study, the designations in parentheses are from \citet{prosser1994} and used in \citet{reipurth2007}.
 The 0535179-051532 system shows both radial velocity variability and a double 
peaked correlation function. From the correlation function, it even appears that we
detect three components in this binary system. The 0535254-053403 system only shows radial
velocity variability. We are not surprised that the 42
others were not detected as spectroscopic binaries because many have wide separations and which would
not yield detectable velocity variations over our time baseline.

\subsection{Kinematic Relation of Stars and Gas}

\begin{figure}
\includegraphics[scale=0.75, angle=-90]{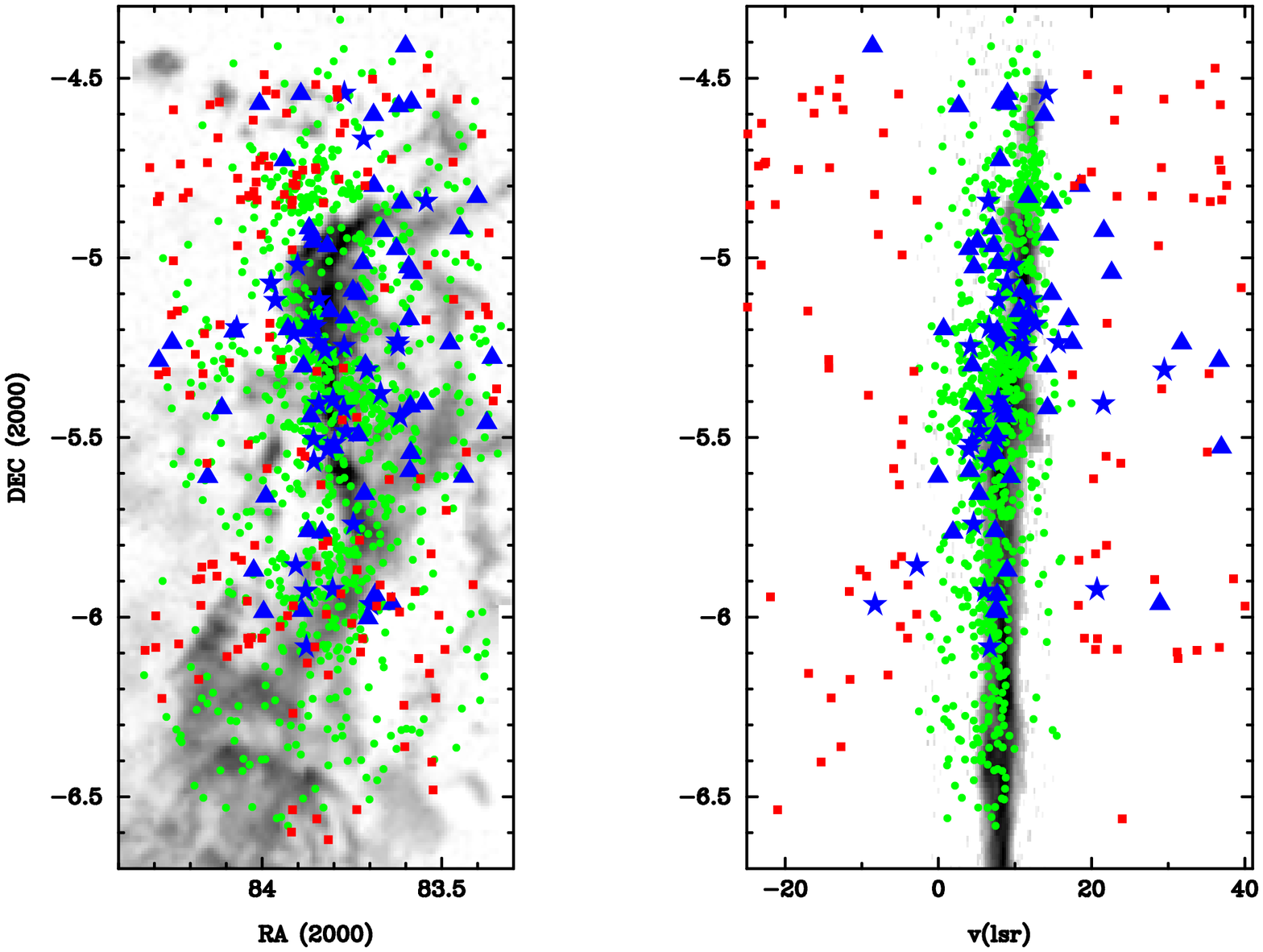}
\caption{Position and Position-Velocity plot of all targets in the ONC using an 
average of all velocity measurements. Binaries are plotted with blue symbols those with an IR excess are plotted as star points,
without are plotted as triangles, and binaries without photometry are plotted as squares. The red squares represent
stars which lie significantly off the cluster velocity, this shows that in general stars with a velocity
deviant from the cluster are located in regions without dense gas.}
\label{pv}
\end{figure}

In Figure \ref{pv}, we plot the spatial distribution of the stars and $^{13}$CO gas in the left panel and 
the position-velocity (PV) relationship of the stars and gas in the right panel.
We have plotted the detected SBs in Figure \ref{pv} as dark blue points, stars for IR excess sources
and triangles for non-excess sources; the non-binary stars are plotted as green circles.
We clearly see that most of the stars are in close association with the
dense filament in the left panel of Fig. \ref{pv} and in the right panel we see that most stars have 
a velocity similar to that of the gas. Additionally, most of the detected
SBs have velocities  similar to that of the gas,
but  note that the velocity of SBs
that are cluster members with multiple observations will average
out to the cluster velocity. Even with many binaries identified, there are still stars that
are clearly not following the gas velocity; these are marked as red squares in Fig. \ref{pv}.
One possible reason for these outliers could be undetected binarity. In total, 154 stars have
vastly different velocities from the gas and only 63 have multi-epoch coverage; thus
 many could still potentially be binaries. We will discuss these outliers further in \S4.

\begin{figure}
\includegraphics[scale=0.75]{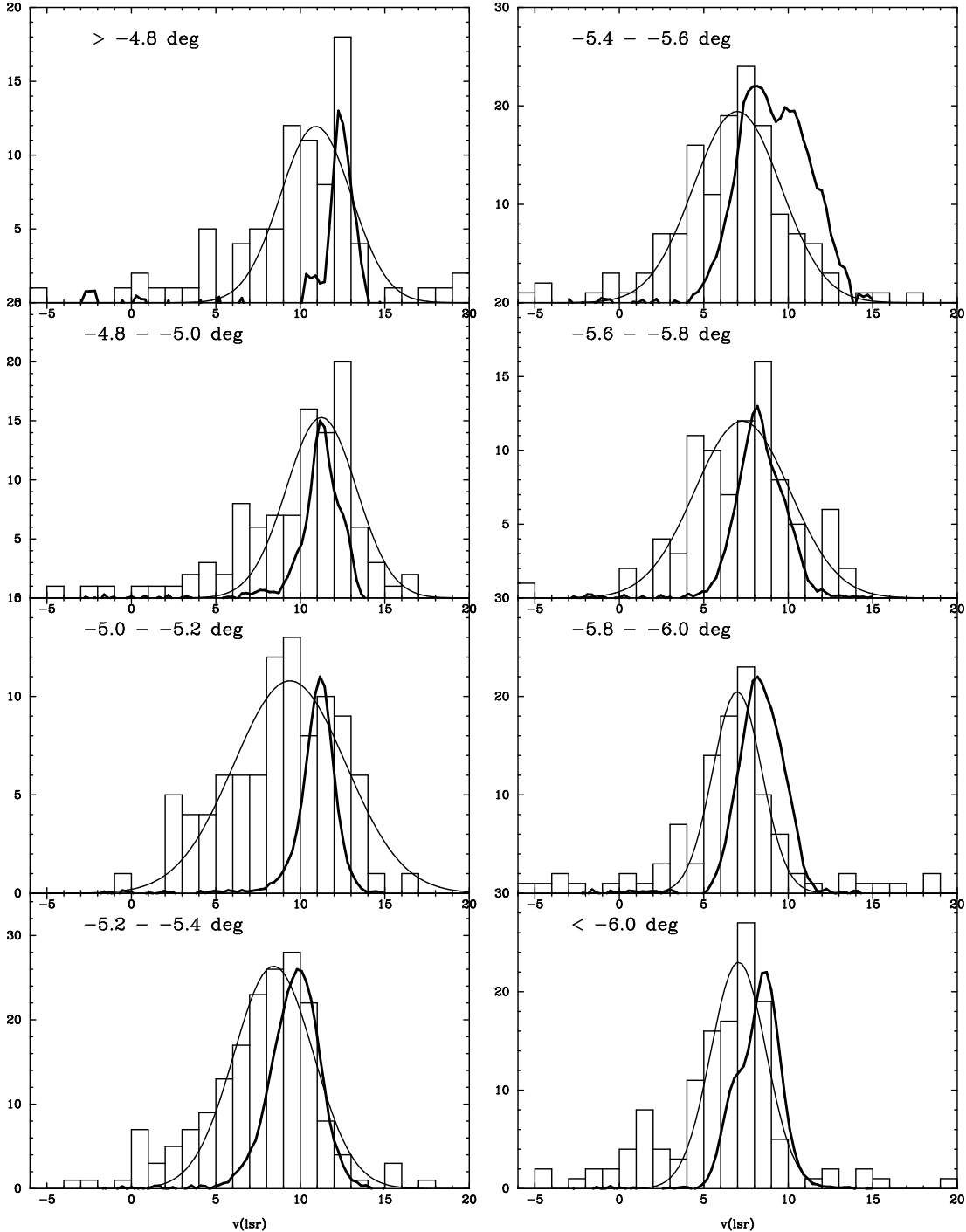}
\caption{Velocity histograms and gaussian fits (thin curves) of non-binary stars in the ONC binned in declination.
 Notice the asymmetry of these distributions
and their velocity tail toward lower velocities. Also, as declination increases, the peak
of the velocity distribution clearly shifts toward higher, red-shifted velocities. The line profiles of $^{13}$CO
are plotted and are within $\sim$1.5 km/s of the peaks in the stellar distribution. The profiles are generally
more narrow that the stellar distributions as velocities with R = 6.0 have $\sim$2 km/s errors.}
\label{histos}
\end{figure}

Looking more closely at the velocity 
structure of the gas and stars, some features stand out.
First, the velocity distributions of the stars in PV space are clearly asymmetric.
In Figure \ref{histos}, we have plotted the velocity histograms of the stars in 
bins of declination along with a gaussian fit to the distribution, though the distributions are 
not exactly gaussian. There is a clear tail toward blue-shifted velocities in some bins 
which is not present toward red-shifted velocities. Then, moving north in declination 
the peaks of the velocity distribution become red-shifted compared to regions south.

We use the gaussian fits, shown in Figure \ref{histos} to estimate the full width at
half-maximum (FWHM) of the velocity distribution in a particular bin.
The FWHMs of the individual bins for gas and stars, as well as the median R, are
listed in Table 13. We have used the median R to estimate
the overall error in radial velocity measurement for a bin and then subtracted
this error in quadrature from the gaussian sigma; the gaussian fits drawn in Figure \ref{histos} are not corrected. 
After correction, we see that most of the ONC has a velocity FWHM between 3.76 and 7.14 km s$^{-1}$. 
 These values compare well to the 1 dimensional FWHM of 5.9 km s$^{-1}$ derived by
\citet{joneswalker1988}. We note that FWHM between -5.0 and -5.2$^{\circ}$ is  larger
than the rest but, in this range of declination lies the OMC-2
and OMC-3 regions. This range has many stars in its tail toward the blue which is making the gaussian fit
overly large. Also, from -5.8$^{\circ}$ and lower, the velocity dispersion is much smaller
 than the rest of the measurements. We will discuss the FWHMs further in \S4.

\begin{figure}
\includegraphics[scale=0.75, angle=-90]{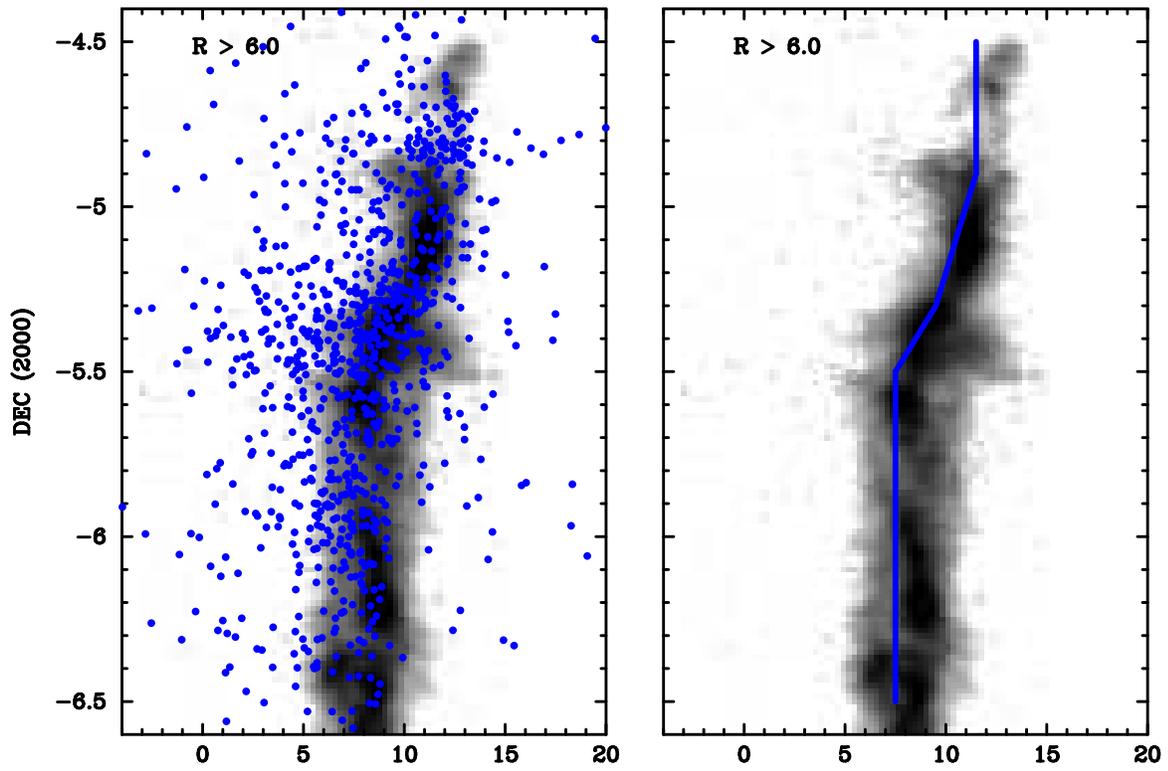}
\caption{Left: Position-Velocity plot of all non-binary targets, using only Hectochelle data with R $>$ 6.0.
Right: Fit to peak of stellar velocity distribution, binned in declination, of the RA range of 84.0 - 83.6$^{\circ}$.
Bins are the same as in Figure 2.}
\label{pvgauss}
\end{figure}

We present a PV plot of a narrow velocity range in the left panel of Figure \ref{pvgauss}
to more closely show the agreement between the stars and gas. In the right panel of Figure \ref{pvgauss},
we have plotted a fit to the peak of the velocity distributions shown in Figure \ref{histos}.
However, to fit the peak we have isolated the targets associated with the main filament from 83.6 to 84.0$^{\circ}$ in RA 
to minimize outlier contributions.
Figure \ref{pvgauss} clearly demonstrates that the gas and stars are well correlated 
within a narrow velocity range over the entire expanse of the molecular cloud.
Notably, both the stars and gas show an abrupt shift toward greater
velocities at a declination of about -5.4$^{\circ}$. This velocity shift is
reflected in the histogram from -5.2 to -5.4$^{\circ}$ in Figure \ref{histos} as
a broad peak in the velocity distribution. The broad peak results from the stellar
velocities closely following the molecular gas velocity through the velocity shift. 
The velocity shift takes place just north of the Trapezium, approximately at the center
of the gaseous filament. The characteristic LSR velocity of the gas  and stars before the shift is about 8 km s$^{-1}$, 
after the shift it is about 11 km s$^{-1}$. 

\section{Discussion}

\subsection{Spectroscopic Binary Population}

The binary fraction of the ONC has garnered much attention recently. The \textit{Hubble Space Telescope}
has made it possible to observe binary stars in the ONC down to $\sim$0.15$^{\prime\prime}$ (60 AU) separations.
The most recent of these studies  find that only $\sim$8.8\% of 
ONC stars are binaries; slightly deficient compared to the field stars and a factor of 2 lower than 
Taurus \citep{reipurth2007}. The leading theory for this  observation is that dynamical interactions in the dense cluster environment
disrupt wide binary  systems \citep{reipurth2007}. In order to test this explanation,
we must determine the binary frequency for all separations. This would enable us to
tell if all binary systems are deficient or if there is a certain separation distance where the ONC becomes 
deficient. However, as the visual searches are not sensitive
to close separations, a large population of binary stars could still be present but only detectable
through radial velocity monitoring. 

\begin{figure}
\includegraphics[scale=0.55]{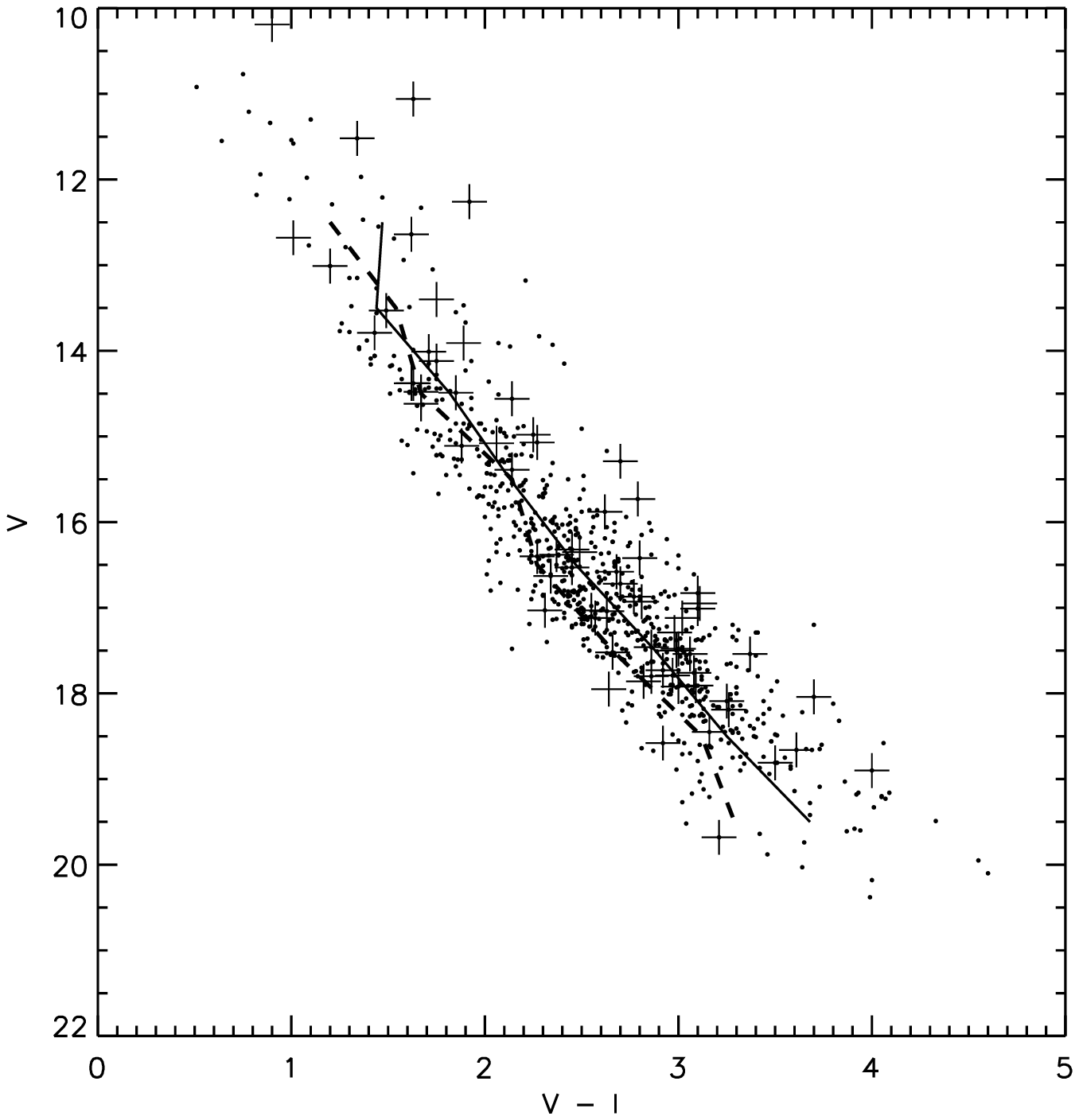}
\includegraphics[scale=0.55]{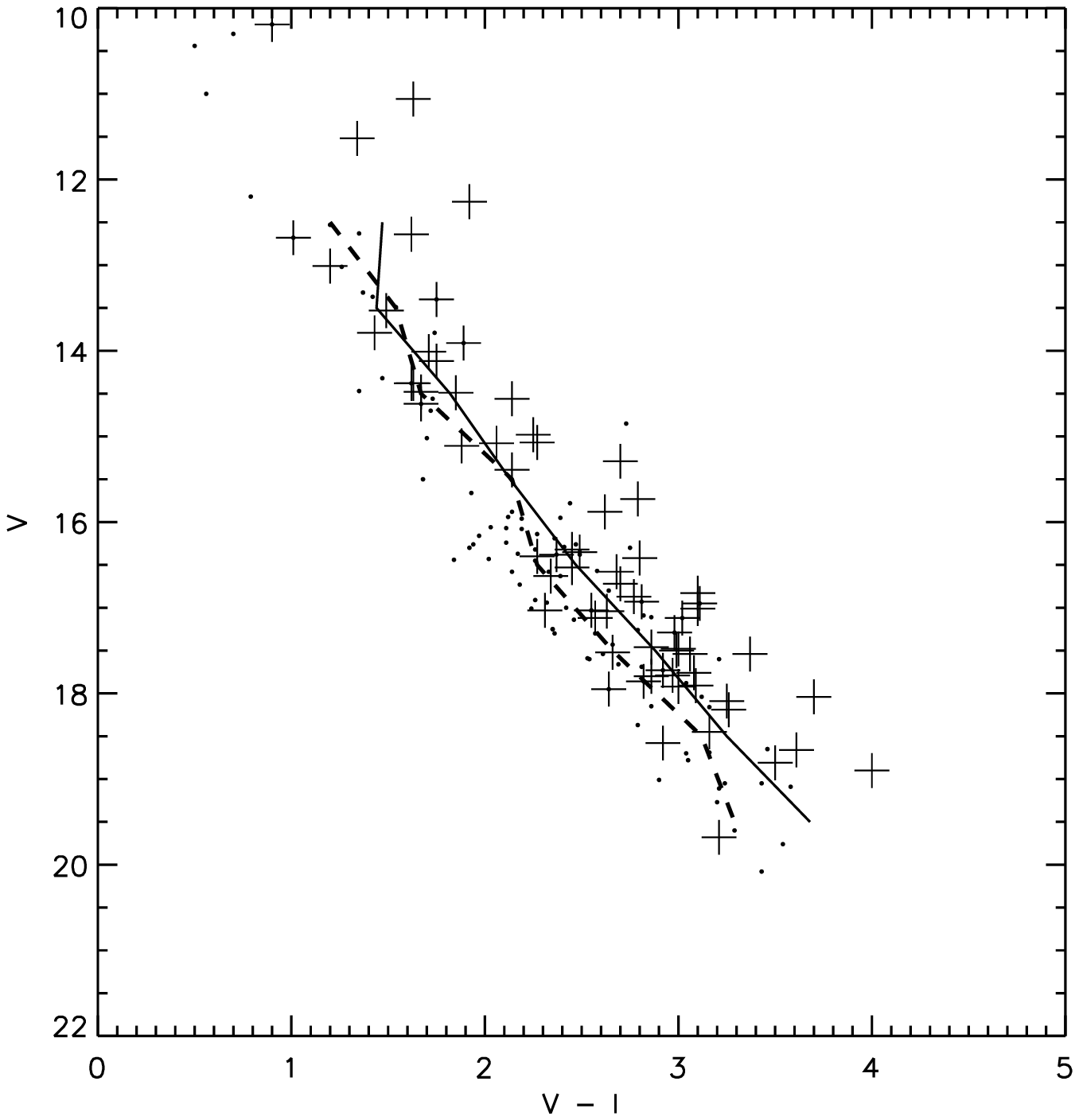}
\caption{Optical color magnitude diagrams of our ONC targets with optical photometry from Rebull et al. (2001) 
and Hillenbrand et al. (1997), the photometry are not corrected for extinction. The left panel is the CMD of stars with radial velocity that is consistent with cluster 
membership, -2.0 $<$ V(lsr) $<$ 18.0. The right panel is the CMD of stars with velocities outside 
the velocity range for cluster membership. The dots are stars in our radial velocity catalog and the plus
 signs overplotted are spectroscopic binaries, we do not separate the binaries based on their average velocity. 
The \textit{solid line} in both panels is the median V-I color for stars with velocities consistent with the ONC,
the \textit{dashed line} is the median V-I color for stars with velocities outside the range for cluster membership. 
The median of the V-I color for both groups of stars show that the stars outside the cluster velocity range are
bluer and fainter than the general ONC suggesting that they are indeed a separate population.}
\label{optcmd}

\end{figure}

Presently, we have identified 89 binaries or 11.5\% of the total sample with multi-epoch coverage.
However, our dataset is not complete enough to yield an accurate final estimate; 
our value of 11.5\% should be regarded as a lower limit. If we used a lower R cutoff
 for our $\chi^2$ routine or lower probability restriction, we would add more binaries to the sample.
 Further observations will likely confirm additional
systems. To characterize the binary systems with respect to the rest of the ONC, we have plotted
the detected binaries as crosses on the V - I CMD of the ONC in Figure \ref{optcmd}. 
Most of the binary stars trace a binary sequence with magnitudes slightly greater than the 
median of the ONC.

\begin{figure}

\plotone{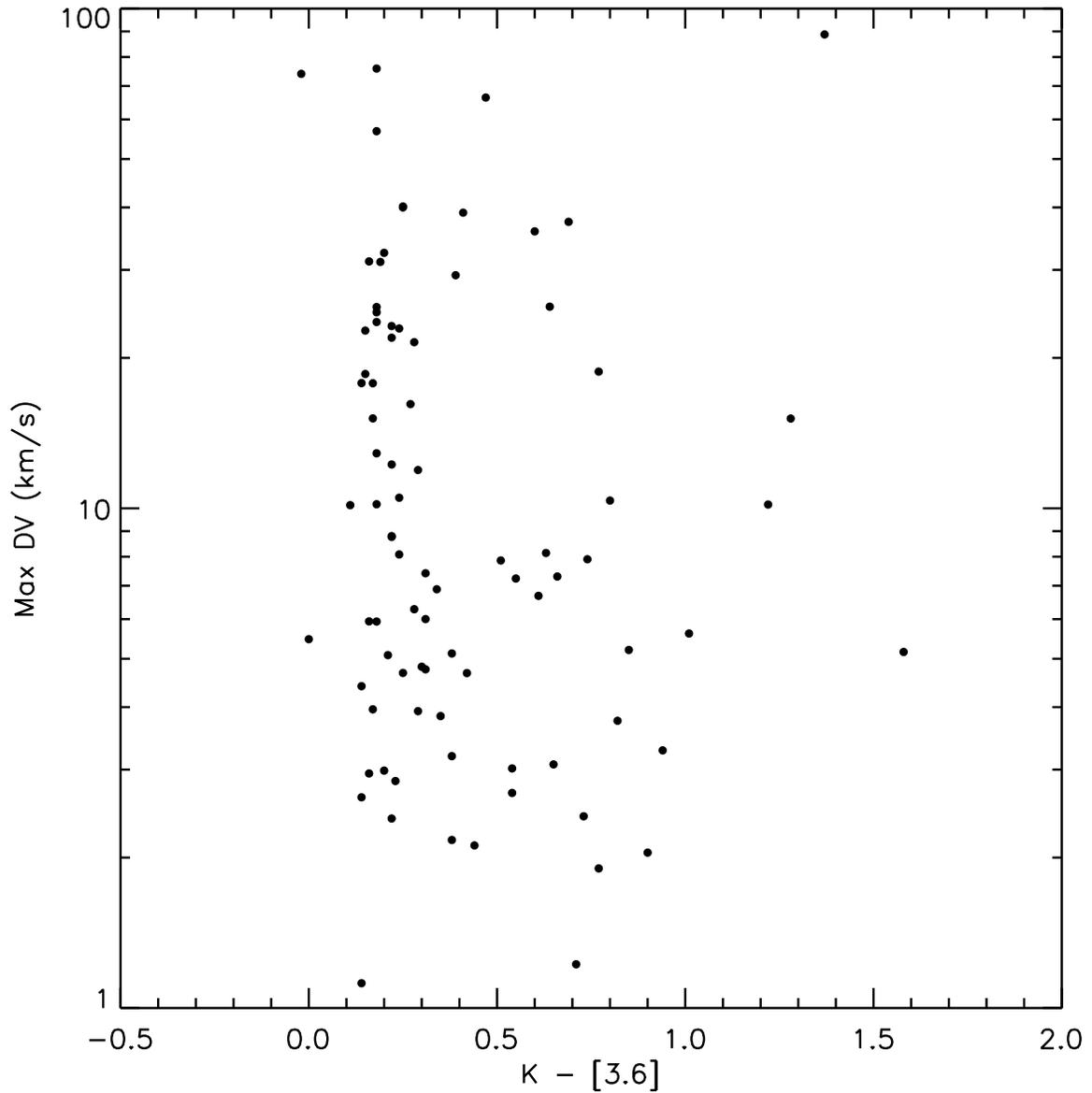}
\caption{Maximum velocity difference from average or difference between correlation peaks for
 probable binaries versus Ks-band - IRAC 3.6$\mu$m color, a value of K - [3.6] $>$ 0.5 indicates
an IR excess. As expected, binary stars without an IR excess are clearly more plentiful than those with excess,
though several stars with high velocity differences do have an IR excess which is not expected.}
\label{maxdvcolor}
\end{figure}

Curiously, we have have found that $\sim$30\% of our detected binary systems also
show an IR excess using the K - IRAC 3.6 $\mu$m  versus IRAC 3.6 - 4.5$\mu$m
color-color diagram indicating the presence of a circumstellar disk. In Figure \ref{pv}, the positions of binary
stars are plotted, and those having an IR excess are marked with star points.
Also, to assess the separation of these systems we plot maximum velocity difference, a surrogate for semi-major axis,
versus K - [3.6] in Figure \ref{maxdvcolor}. The maximum velocity difference is determined by measuring the difference 
between correlation peaks for the SB2s and simply the maximum radial velocity minus the minimum for systems detected
by velocity variability.
We see that some binaries with near-IR excesses at 3.6 $\mu$m also have velocity variations larger than 10 km s$^{-1}$, 
which is $\sim$10AU for a system of 1 M$_{\sun}$ total mass.
The majority of stars with velocity differences $>$ 10 km s$^{-1}$ are binaries identified from
the double-peaked correlation function.

The significant number of spectroscopic binary systems with near-IR excess is
surprising. For most binary systems
compact enough to be detected spectroscopically, the companion star is expected to have evacuated
the inner disk as in Coku Tau/4 \citep{ireland2008}. Thus the binaries without a near-IR excess
could be transition objects or have an inner disk hole, these objects would then show an IR excess long-ward of 10$\mu$m.
 Seven others have near-IR excesses but have velocity variations less than 10 km s$^{-1}$. These systems may be wider binaries
with a truncated outer disk.  Perhaps there is an upper limit to the eccentricity of the companion
orbit in order to retain the inner disk. This appears to be the case of GG Tau \citep{mccabe2002} which
is shown to have the spectrum of a normal disk \citep{furlan2006} and an eccentricity of $\sim$0.3. It is also
possible that the disk clearing by the companion has simply not been completed in these relatively young systems.
Mid-infrared spectra and MIPS photometry in forthcoming studies from \textit{Spitzer} and
further radial velocity monitoring will constrain the binary orbits and shed light on the disk
structure of these young systems.

\subsection{Velocity Dispersions}

In \S3, we presented the FWHMs of the ONC binned in declination. We noted that the two most 
southern and most northern declination bins in Figure \ref{histos} have a much lower FWHMs
than the rest of the cloud. The FWHM seem to follow the
trend of being greater where there is moderately dense gas present, and smaller where there is little
gas present or the gas has been evacuated, specifically south 
of the Orion Nebula and in the NGC 1977 region (Figure \ref{upper}). The values for the FWHMs
of the stars and gas are given in Table 13.

However, looking at the $^{13}$CO linewidths in Figure \ref{histos}, we do not observe the
trend found in the stellar velocity distributions. Even after correction for measurement error
in the radial velocities, most linewidths are more narrow than the stellar
 velocity distributions by $\sim$2 km s$^{-1}$, except in the two most southern regions. Also, there is a 
trend of larger linewidths south of the Trapezium and smaller linewidths north. The differences
between the stellar velocity distribution and the gas line profiles may be due to effects such as
unresolved binary motion, gas dispersal, stellar interactions, and/or underestimates of associated
errors.

We do notice that not all line profiles are symmetric, though we are making these plots of the
$^{13}$CO by averaging over a large area (0.2$^{\circ}$ (Dec.)x 0.5$^{\circ}$ (RA)). 
Thus in the north the line emission may be dominated by the filament which is kinematically coherent.
While in the south, the gas is less dense and there seems to be multiple velocity components in the
lines. In the -5.4$^{\circ}$ to -5.6$^{\circ}$ bin, there are clearly two velocity components 
which are probably due to the Trapezium stars blowing the gas away producing a red and blue-shifted 
component. We can see this structure in more detail
in Figure \ref{pvgauss}, where there is a wide velocity spread in the gas at -5.5$^{\circ}$ and 
there is a decrement of intensity compared to regions directly north and south.

\subsection{Asymmetry and Outliers in the Velocity Distribution}

Thus far, we have discussed how closely the stars follow the gas as shown in Figure \ref{pvgauss};
however, we clearly see more stars blue-shifted from the gas than are red-shifted.
Ordinarily, we would expect there to be an even distribution  of blue and red-shifted stars. Some of this asymmetry
may be due to a global zeropoint offset between the stellar velocities and the gas velocity.
Shown in Figure \ref{histos}, in some regions the velocity distributions of the stars and gas
can differ by about 1 - 1.5 km s$^{-1}$. However, a velocity offset between the stars and gas would only
account for some of the asymmetry. 

There are a few other possibilities that could explain the blue-side asymmetry. One is that the
effect could be systematic in nature. Incomplete sky subtraction could cause a slight blue-shift
since the heliocentric corrections were always blue-ward of the cluster velocity. Another possibility
is that we are detecting a fainter foreground population, though the CMD of stars these slightly
blue-shifted stars is not different from the rest of the cluster. In all, we do not know the exact cause of the shift, but its presence
does not take away from our results.

In addition to the asymmetry, in the left panel of Figure \ref{pv} we see 
that some stars have velocities more than 10 km s$^{-1}$ different from the cluster velocity, 156 in total.
We acknowledged in \S3 that some of these may be binaries as our multi-epoch coverage
is not complete; though, it is unlikely for most of these to be binaries.
To examine the spatial relationship of the outliers to the rest of the cluster, the targets with outlying velocities
have been plotted as red squares on Figure \ref{pv}. Many of these deviant velocity targets are 
located in regions without much $^{13}$CO emission on the edges of the survey area and there seems to be zone of avoidance
around the dense filament. This implies that most of these additional stars selected to fill
more fibers are probably foreground sources and not physically associated with the ONC.

We have plotted the sources with 
V and I-band photometry on the optical CMD in Figure \ref{optcmd} and we have plotted
the median V - I color for both the stars within 10 km s$^{-1}$ of the gas and those outside 10 km s$^{-1}$.
We see that the two populations are clearly separated by $\sim$0.2 mag in V - I and $\sim$1 mag V.
This separation in the optical CMD suggest and large velocities together strongly suggest
that these stars are older and not members of the ONC. The inclusion of stars such as these and binaries
in studies of the ONC could be responsible for the wide age
 spread quoted in some studies \citep[e.g.][]{jeffries2008,tkm2006}. Some of these
outlying stars do have IR excesses detected with IRAC photometry; however, we do not have enough observations to 
confirm that they are not spectroscopic binaries. If these IR excess stars are not binaries, then they may have been
 dynamically ejected from the ONC, and this would mean that circumstellar disks can survive a strong dynamic interaction.
More simply, they could be a members of the older foreground population whose disks have persisted for longer than
expected.

\subsection{Effects of Stellar Feedback}
\begin{figure}
\begin{center}
\includegraphics[scale=0.8, angle=-90]{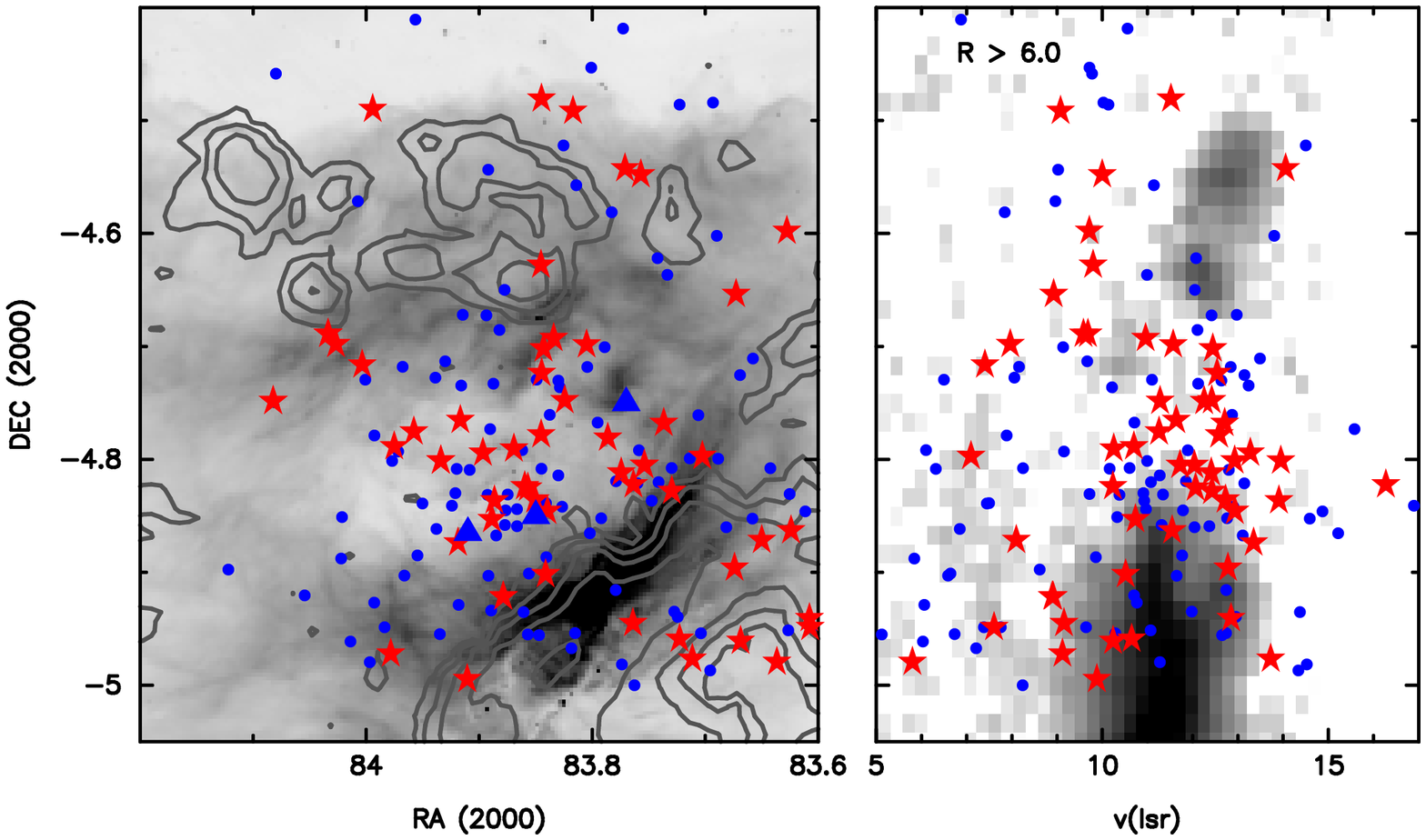}
\includegraphics[scale=0.33, angle=-90]{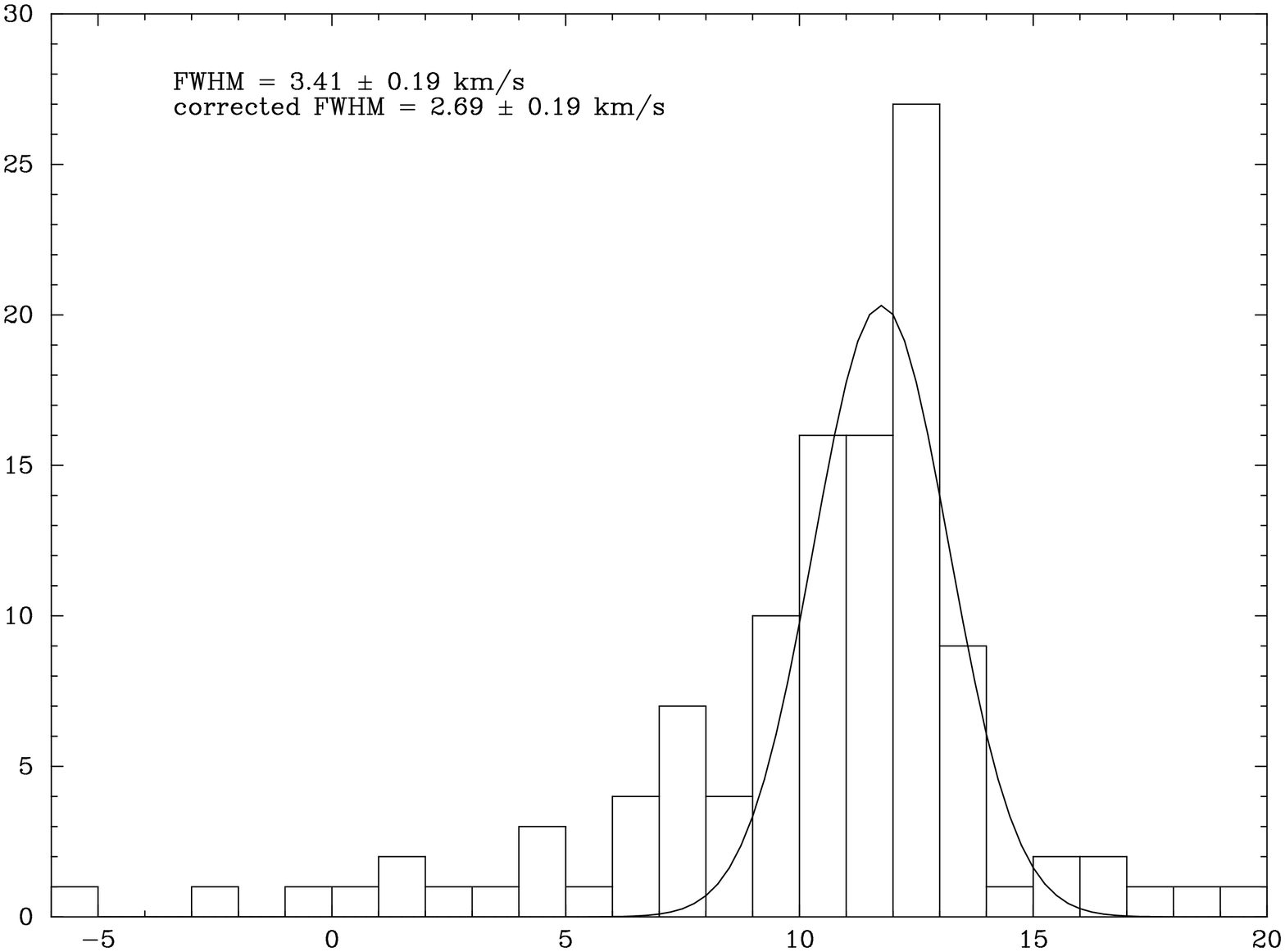}

\end{center}
\caption{Upper Left: Position plots stars with 5.0 $\ge$ V(LSR) $\ge$ 20.0, the contours are the $^{13}$CO emission.
Upper Right: Position-velocity plot of the region shown in the upper left. Lower Panel: Velocity histogram and
gaussian fit for the stars within the RA range of 83.6 - 84.1$^{\circ}$ and -4.5 - -4.9$^{\circ}$ in declination.
Note that there is no dense gas where we see the densest concentration IR excess stars (stars) and
non-excess stars (circles) in the center of the PAH ring. The B stars are marked as large triangles, the most massive
star is spectral type B1V at $\alpha$=83.85 $\delta$=-4.85 in the image, this is also the most likely ionization source
for the very bright PAH feature directly to its south. The other massive stars are spectra type B3V. Also, the stars within
this region have a very tight velocity distribution (lower panel).}
\label{upper}

\end{figure}

With all the massive stars in the ONC, we should be able to directly observe
its effects on the kinematic structure. As discussed in \S4.2,
the gas at spatial locations near the Trapezium stars has a very wide spread in
velocity as evidenced in Figures \ref{histos} \& \ref{pvgauss}, meaning that it is 
being blown away. However, north of the Trapezium in NGC 1977,
we see a region that has been evacuated of dense gas and a "mini cluster" of stars is left.
The most powerful star in NGC 1977 is spectral type B1V (HD 37018) and 
there are also two B3V stars present (HD 37077 \& HD 36958). These stars are fairly weak in comparison to the 
Trapezium, yet they have nearly evacuated all dense gas within a 1 pc radius. This is a clear demonstration 
that B-stars rapidly disperse gas; this effect is also seen in the clusters IC348, NGC7129, and IC5146
 \citep[][and references therein.]{allen2007}.  The \textit{Spitzer} image 
in the left panel of Figure \ref{upper} shows a ring of emission, probably due to Polycyclic Aromatic Hydrocarbons (PAHs),
 around an evacuated central cluster region \citep{peterson2008}. This ring was also
shown in the MSX images of \citet{kraemer2003} but not discussed.

Nearly all the stars located inside the PAH ring are located in a very tight velocity
group; the stars preferentially occupy the 10 - 15 km s$^{-1}$ velocity range, see Figure \ref{upper}.
This group has one of the tightest and most symmetrical velocity distributions in our data. 
The tight velocity distribution is likely due to the lower mass enclosed by this small cluster due to 
the evacuation of the gas. The stars with higher velocities probably escaped and may occupy the 
red and blue-shifted velocity tails in Figure \ref{histos}. Also, even though most gas is gone, the stars 
still tightly correlate with the velocity of the gas to the north and south.

 There is also some direct evidence 
for forced ejection of material from this region, in the right panel of Figure \ref{upper} there is a faint clump
at -4.7$^{\circ}$ with a blue shifted velocity compared to the stars. Also, in Figure \ref{pvgauss}, where
the PV plot is taken over a wider range in RA, there is a extension of the gas to blue-shifted velocities
at  -4.85$^{\circ}$. The gas dispersal 
must have happened quickly compared to the timescale of disk dissipation as 70\% 
of stars in this regions have an IR excess (Megeath et al. in prep.). Though, because we are bias toward optical stars, only
36\% of our surveyed stars have an IR excess.

\subsection{Global Kinematic Structure}

Our dataset of precise stellar radial velocities, in conjunction with the detailed gas kinematics, provide
an unprecedented opportunity to study the dynamics of a cluster whose structure may still reflect the
initial conditions of formation. In
\S3, we roughly outlined how the dense $^{13}$CO gas and stars occupy the same PV space. The regularity
of the velocity structure is quite astounding. In terms of physical sizes, the velocity of the gas and
stars correlate well over a 16 pc projected length. However, the northern and southern parts of the cluster have
distinctly different characteristic velocities. They are connected by the striking velocity gradient just north
of the Orion nebula (OMC-2/3 regions). In a 2 pc (0.25$^{\circ}$) region, the stars and gas shift toward the 
red by 2.5 km s$^{-1}$. It is important to point out that there is not a corresponding velocity gradient
in the southern part of the cluster.

We posit that the velocity gradient in the OMC-2/3 region and distinct kinematics of the north and south 
regions of the cluster are signatures of large-scale infall. Both the gas and stars seem to be falling
in from the north; this is strong evidence that the system is not in dynamical equilibrium. 
Gravitational collapse presents the simplest explanation for the observed global velocity coherence
and why the stars so closely follow the velocity gradient. 
This condition would of course require that the
northern end of the Orion A cloud is somewhat closer than the southern end of the cloud,
which is possible given the uncertainties in the 3D structure of the ONC and Orion A cloud.
Infall actively taking place in the ONC is very strong
evidence for its youth ($\sim$ 1 crossing time, 1 Myr), along with estimated stellar ages \citep{hill1997}.

Our idea of large-scale infall is supported by the N-body simulations of \cite{proszkow2008}. 
These simulations modeled clusters with parameters similar to the ONC, elongated in the north-south direction and comparable in
total mass. They found that a velocity gradient, similar to what is observed in the ONC, could only be present
if the the initial conditions of the cluster were sub-virial. If a cluster was initially in virial
equilibrium, only a very slight velocity gradient is witnessed. In addition, the magnitude of velocity gradient
observed in the model, is highest at the first crossing time. Thus, these simulations, in addition to our kinematic data, are strong
evidence that the ONC is a cluster which formed in sub-virial initial conditions and is currently about 1 crossing 
time in age.

While we argue that sub-virial collapse is the most appealing explanation, 
several studies have focused on modeling the ONC as a quasi-equilibrium
system \citep[e.g.][]{tkm2006, scm2005,huff2007}.
However, the velocity gradient we observe from the north to the south should not
be present if the ONC formed in a quasi-equilibrium state. Similarly, we argue that the
ONC cannot be much more than 1 crossing time old because if the cloud has been
evolving over several dynamical times, the gas will have shocked and 
the velocity structure should have been washed away. 
The studies cited above also assume that the stellar population of the central ONC is relaxed.
However, it is uncertain if the central ONC is relaxed because the stars just north of the Trapezium
follow the velocity gradient. But the there is some evidence for the stars to be undergoing
violent relaxation \citep{fiegelson2005}.

Other alternatives could be that a supernova north of the ONC
could have induced the shift. However, it is unclear why the velocity of the southern
part of the cloud was not shifted as well.  
Also, large scale rotation of the cloud has been proposed by \citet{kutner1977}
but this possibility does not seem likely because the cloud would be rotating near breakup.
Also, stellar feedback could give rise to
the velocity gradient, but stellar feedback blows the gas away from the stars. If the 
velocity gradient was induced \textit{a posteriori} of star formation, the stars would not should
not follow the velocity gradient, which they clearly do.

\section{Summary}
We have presented new radial velocity measurements for nearly 1124 stars in the ONC, building on the work
of \citet{furesz2008}. With these additional kinematic data, we have refined our determination
of the star to gas kinematic relationship in the ONC. The accuracy of the radial velocity
measurements has enabled us to clearly show that the stars closely follow the kinematic
structure of the moderate density, $^{13}$CO gas. 

We have begun an initial characterization of the spectroscopic binary population of the ONC.
We have obtained multi-epoch observations for over half of our total sample and find 89 (11.5\%)
of the stars to be spectroscopic binaries. However, this number is currently a lower
limit at best as we are focusing on detection rather than detailed study of these binaries.
We are continuing our radial velocity monitoring program to further
identify spectroscopic binary stars in the ONC. In the future, a more directed program is necessary
to constrain the orbital parameters of these spectroscopic binaries.

North of the central ONC in NGC 1977, all gas has been evacuated within about $\sim$1 pc of
the most luminous star. The stars within this evacuated region occupy a very tight velocity range,
possibly indicating that higher velocity stars have escaped after the dispersal of the gas.
About 70 \% of the stars show an IR excess, thus we can infer that the gas must have been 
evacuated quickly. This observation shows that even the wimpiest of high mass stars can clear out their surrounding molecular
material quickly.

Overall, the most striking feature in the kinematic structure is the sharp velocity shift 
($\sim$2.5 km s$^{-1}$) toward the red just north of the Trapezium. We argue that this velocity gradient is a signature 
of large-scale infall; the gas and stars in the OMC-2/3 and NGC 1977 regions have a red-shifted
 velocity compared to areas farther south and the gradient is traced by both the 
gas and stars. These observational
results taken with the results of recent N-body simulations are consistent with the view of
the ONC as a collapsing, sub-virial cluster \citep{proszkow2008}.

The authors wish to thank the staff of the MMT and Magellan telescopes
and the Hectochelle queue observers of Fall 2008 for their efforts in
obtaining the data used in this paper. We thank the referee I. Bonnell
for helpful comments which helped improve the clarity of this paper.
 We also thank M. Walker and E. Olszewski for
assistance with MIKE Fibers data acquisition/reduction and 
Jesus Hernandez for useful discussions. J. T. and L. H. acknowledge funding
NASA grant 1342979.

Facilities: MMT, Magellan, Spitzer
\bibliographystyle{apj}
\bibliography{ms}

\clearpage
\input{tab1}
\input{tab2}
\input{tab3}
\input{tab4}
\input{tab5}
\input{tab6}
\input{tab7}
\input{tab8}
\input{tab9}
\input{tab10}
\input{tab11}
\input{tab12}
\input{tab13}

\end{document}

%% file: tab1.tex
\begin{deluxetable}{lllllllllll}

\tablewidth{0pt}

\tabletypesize{\scriptsize}
\tablecaption{MIKE Fibers Observations}
\tablehead{
  \colhead{Field ID} & \colhead{Start Date} & \colhead{Julian Date} &  \colhead{RA} & \colhead{Dec.} & \colhead{Airmass} & \colhead{Exposure Time} & \colhead{Binning}  & \colhead{Filter} & \colhead{Zeropoint Shift}\tablenotemark{a} & \colhead{Number of Targets}\\
                 & \colhead{UT Date} & \colhead{(2454000)} &  \colhead{(J2000)} & \colhead{(J2000)} & & \colhead{\# $\times$ seconds} & & & \colhead{km s$^{-1}$}\\

   }
\startdata
A-1 & 01-20-2007 & 120.579 & 05:35:21.4   & -05:57:38.8  & 1.101 & 3x1200 & 2x2 & Mg & -5.72\tablenotemark{b} $\pm$ 0.4, 1.98\tablenotemark{c} $\pm$ 0.23 & 252\\
B-1 & 01-21-2007 & 121.560 & 05:34:59.7   & -05:25:42.2  & 1.136 & 3x1200 & 2x2 & Mg &  0.54 $\pm$ 0.5 & 250\\
C-1 & 01-24-2007 & 124.560 & 05:35:16.40  & -05:15:01.4  & 1.123 & 3x1200 & 2x2 & Mg & -1.02 $\pm$ 0.4 & 227\\
A-2 & 03-20-2008 & 555.517 & 05:35:21.4   & -05:57:38.8  & 1.339 & 4x1200 & 2x2 & Mg & -4.16\tablenotemark{b}$\pm$ 0.38, -3.13\tablenotemark{c}$\pm$ 0.23 & 252\\
B-2 & 03-27-2008 & 552.550 & 05:34:59.7   & -05:25:42.2  & 1.500 & 2x1200 & 2x2 & Mg & -1.17 $\pm$ 0.6 & 250\\
C-2 & 03-31-2008 & 555.515 & 05:35:16.40  & -05:15:01.4  & 1.279 & 3x1200 & 2x2 & Mg & -1.17 $\pm$ 0.5& 217\\
C-3 & 04-03-2008 & 561.488 & 05:35:16.40  & -05:15:01.4  & 1.288 & 3x1200 & 2x2 & Mg & -1.25 $\pm$ 0.5& 227\\
D-2 & 03-28-2008 & 553.539 & 05:35:29.7   & -04:50:00.80 & 1.466 & 3x1200 & 2x2 & Mg & 0.81 $\pm$ 0.6& 231\\

\enddata
\tablenotetext{a}{Shift applied is relative to the velocities in \citet{furesz2008}.}
\tablenotetext{b}{Shift was applied only to spectra observed with the red channel of the MIKE spectrograph.}
\tablenotetext{c}{Shift was applied only to spectra observed with the blue channel of the MIKE spectrograph.}
\tablecomments{We only had enough good velocities for Field A to treat the  blue and red
spectrograph channels independently.} 
\end{deluxetable}

%% file: tab2.tex
\begin{deluxetable}{lllllllllll}

\tablewidth{0pt}

\tabletypesize{\scriptsize}
\tablecaption{Hectochelle Observations}
\tablehead{
  \colhead{Field ID} & \colhead{Date} & \colhead{Julian Date} &  \colhead{RA} & \colhead{Dec.} & \colhead{Airmass} & \colhead{Exposure Time} & \colhead{Binning}  & \colhead{Filter} & \colhead{Zeropoint Shift}\tablenotemark{a} & \colhead{Number of Targets}\\
                 & \colhead{UT Date} & \colhead{(2454000)} &  \colhead{(J2000)} & \colhead{(J2000)} & & \colhead{\# $\times$ seconds} & & & \colhead{km s$^{-1}$}
}
\startdata
F1-E1 & 10-25-2007 & 399.000 & 05:35:13.87  & -04:50:15.37 & 1.317 & 3x1200 & 2x2 & RV31 & 0.45 $\pm$ 0.07 & 192\\
F1-E2 & 10-27-2007 & 400.893 & 05:35:13.87  & -04:50:15.37 & 1.310 & 3x1200 & 2x2 & RV31 & 1.43 $\pm$ 0.04 & 192\\
F2-E1 & 10-27-2007 & 401.003 & 05:35:12.75  & -05:19:42.90 & 1.355 & 3x1200 & 2x2 & RV31 & 1.40 $\pm$ 0.13 & 203\\
F2-E2 & 10-28-2007 & 401.836 & 05:35:12.75  & -05:19:42.90 & 1.602 & 3x1200 & 2x2 & RV31 & 1.63 $\pm$ 0.11  & 203 \\
F3-E1 & 10-29-2007 & 402.876 & 05:35:05.01  & -05:26:01.25 & 1.358 & 3x1200 & 2x2 & RV31 & 1.34 $\pm$ 0.24  & 202\\
\enddata
\tablenotetext{a}{Shift applied is relative to the velocities in \citet{furesz2008}.}
\end{deluxetable}

%% file: tab3.tex
\begin{deluxetable}{lllllllll}
\rotate
\tablewidth{0pt}

\tabletypesize{\scriptsize}
\tablecaption{Velocity Summary}
\tablehead{
  \colhead{2massID} & \colhead{RA} & \colhead{Dec.} &  \colhead{$\overline{RV}$\tablenotemark{a}} & \colhead{N\_obs\tablenotemark{b}} & \colhead{V} & \colhead{V - I} & \colhead{Multiplicity} & \colhead{Field ID\tablenotemark{c}}\\
  & \colhead{(J2000)} & \colhead{(J2000)} & \colhead{(km s$^{-1}$)} \\

   }
\startdata
0533179-052138 & 05:33:17.957 & -05:21:38.63 & 0.0 $\pm$ 0.0 &        0 &  18.27 &  3.29 &    1 & F2-E1, F2-E2, F22                                       \\ 
0533204-051123 & 05:33:20.436 & -05:11:23.97 & 24.3 $\pm$ 0.5 &        1 &  17.61 &  3.09 &    1 & F21                                                     \\ 
0533225-053240 & 05:33:22.581 & -05:32:40.07 & 23.9 $\pm$ 1.0 &        2 &  16.9  &  2.51 &    1 & F3-E1, F22                                              \\ 
0533233-052153 & 05:33:23.321 & -05:21:53.11 & 47.2 $\pm$ 0.1 &        3 &  16.06 &  2.03 &    1 & F2-E1, F2-E2, F22                                       \\ 
0533234-044234 & 05:33:23.492 & -04:42:34.44 & 0.0 $\pm$ 0.0 &        0 &   0    &  0    &    1 & F11                                                     \\ 
\enddata
\tablenotetext{a}{Velocities from \citet{furesz2008} are included in average. For a measurement to be included in the average R must be $>$ 6.0.}
\tablenotetext{b}{Number of observations with R $>$ 6.0.}
\tablenotetext{c}{F11, F21, F22, F31, S3, S2, S1 correspond to observations presented in \citet{furesz2008}.}
\end{deluxetable}

%% file: tab4.tex
\begin{deluxetable}{lllll}
\rotate
\tablewidth{0pt}

\tabletypesize{\scriptsize}
\tablecaption{Hectochelle Field 1 Velocity Data}
\tablehead{
  \colhead{2massID} &  \colhead{RV F1-E1} & \colhead{RV F1-E2} & \colhead{R F1-E1} &\colhead{R F1-E2}  \\
    \colhead{}    &  \colhead{(km s$^{-1}$)} & \colhead{(km s$^{-1}$)} & & \\

   }
\startdata
0533333-043918 & -6.9 $\pm$ 0.2  & -6.3 $\pm$ 0.2  & 41.8 & 37.1 \\ 
0533364-044949 & 21.2 $\pm$ 1.8  & 27.6 $\pm$ 1.8  & 9.3 & 8.5 \\ 
0533377-043349 & 26.3 $\pm$ 0.3  & 25.6 $\pm$ 0.4  & 28.2 & 27.4 \\ 
0533391-043807 & 27.5 $\pm$ 0.7  & 29.6 $\pm$ 0.8  & 10.5 & 9.3 \\ 
0533431-044714 & 25.0 $\pm$ 0.9  & 25.2 $\pm$ 0.7  & 9.0 & 10.6 \\ 
\enddata
\end{deluxetable}

%% file: tab5.tex
\begin{deluxetable}{lllll}
\rotate
\tablewidth{0pt}

\tabletypesize{\scriptsize}
\tablecaption{Hectochelle Field 2 Velocity Data}
\tablehead{
  \colhead{2massID} &  \colhead{RV F2-E1} & \colhead{RV F2-E2} & \colhead{R F2-E1} &\colhead{R F2-E2}  \\
    \colhead{}    &  \colhead{(km s$^{-1}$)} & \colhead{(km s$^{-1}$)} & & \\

   }
\startdata
0533233-052153 & 47.2 $\pm$ 0.4  & 47.2 $\pm$ 0.4  & 20.1 & 24.1 \\ 
0533263-051640 & 67.6 $\pm$ 0.6  & 49.5 $\pm$ 2.5  & 12.1 & 9.4 \\ 
0533289-050930 & 106.8 $\pm$ 0.3  & 106.3 $\pm$ 0.2  & 31.4 & 39.6 \\ 
0533293-050749 & 32.6 $\pm$ 0.8  & 31.4 $\pm$ 0.8  & 11.8 & 13.3 \\ 
0533357-050923 & 30.6 $\pm$ 3.4  & 31.0 $\pm$ 2.5  & 9.3 & 10.6 \\ 
\enddata
\end{deluxetable}

%% file: tab6.tex
\begin{deluxetable}{lll}
\rotate
\tablewidth{0pt}

\tabletypesize{\scriptsize}
\tablecaption{Hectochelle Field 3 Velocity Data}
\tablehead{
  \colhead{2massID} &  \colhead{RV F3-E1}  & \colhead{R F1-E1}   \\
    \colhead{}    &  \colhead{(km s$^{-1}$)} &  \\

   }
\startdata
0533225-053240 & 24.7 $\pm$ 0.9  & 8.2  \\ 
0533256-052354 & 70.3 $\pm$ 0.4  & 22.8  \\ 
0533287-052610 & 30.4 $\pm$ 1.6  & 5.6  \\ 
0533298-052735 & -17.5 $\pm$ 0.5  & 18.3  \\ 
0533301-052257 & 22.6 $\pm$ 0.6  & 12.6  \\ 
\enddata
\end{deluxetable}

%% file: tab7.tex
\begin{deluxetable}{lllll}
\rotate
\tablewidth{0pt}

\tabletypesize{\scriptsize}
\tablecaption{MIKE Field A Velocity Data}
\tablehead{
  \colhead{2massID} &  \colhead{RV A-1} & \colhead{RV A-2} & \colhead{R A-1} &\colhead{R A-2}  \\
    \colhead{}    &  \colhead{(km s$^{-1}$)} & \colhead{(km s$^{-1}$)} & & \\

   }
\startdata
0534330-055747 & -2.4 $\pm$ 3.6  & 59.8 $\pm$ 1.9  & 4.6 & 9.4 \\ 
0534338-055638 & -7.4 $\pm$ 2.5  & -3.8 $\pm$ 1.4  & 6.9 & 13.6 \\ 
0534347-055308 & 0.0 $\pm$ 0.0  & 112.3 $\pm$ 4.1  & 0.0 & 3.9 \\ 
0534351-055815 & 23.0 $\pm$ 2.6  & 25.1 $\pm$ 2.7  & 6.9 & 6.4 \\ 
0534377-060233 & 26.4 $\pm$ 1.5  & 24.1 $\pm$ 1.0  & 12.4 & 20.3 \\ 
\enddata
\end{deluxetable}

%% file: tab8.tex
\begin{deluxetable}{lllll}
\rotate
\tablewidth{0pt}

\tabletypesize{\scriptsize}
\tablecaption{MIKE Field B Velocity Data}
\tablehead{
  \colhead{2massID} &  \colhead{RV B-1} & \colhead{RV B-2} & \colhead{R B-1} &\colhead{R B-2}  \\
    \colhead{}    &  \colhead{(km s$^{-1}$)} & \colhead{(km s$^{-1}$)} & & \\

   }
\startdata
0534174-053158 & 0.0 $\pm$ 0.0  & 26.2 $\pm$ 4.3  & 0.0 & 3.7 \\ 
0534181-052833 & 6.3 $\pm$ 4.5  & 0.0 $\pm$ 0.0  & 3.5 & 0.0 \\ 
0534195-053019 & 5.1 $\pm$ 4.4  & 14.0 $\pm$ 3.2  & 3.5 & 5.5 \\ 
0534207-053235 & 29.1 $\pm$ 3.9  & 0.0 $\pm$ 0.0  & 4.2 & 0.0 \\ 
0534209-052448 & 0.0 $\pm$ 0.0  & 33.8 $\pm$ 3.4  & 0.0 & 5.0 \\ 
\enddata
\end{deluxetable}

%% file: tab9.tex
\begin{deluxetable}{lllllll}
\rotate
\tablewidth{0pt}

\tabletypesize{\scriptsize}
\tablecaption{MIKE Field C Velocity Data}
\tablehead{
  \colhead{2massID} &  \colhead{RV C-1} & \colhead{RV C-2} & \colhead{RV C-3} & \colhead{R C-1} & \colhead{R C-2} & \colhead{R C-3} \\
    \colhead{}    &  \colhead{(km s$^{-1}$)} & \colhead{(km s$^{-1}$)} & \colhead{(km s$^{-1}$)}& & & \\

   }
\startdata
0534269-051803 & 0.0 $\pm$ 0.0  & 0.0 $\pm$ 0.0  & -24.5 $\pm$ 4.4  & 0.0 & 0.0 & 3.6\\ 
0534290-051414 & 25.7 $\pm$ 1.9  & 20.3 $\pm$ 2.0  & 21.4 $\pm$ 2.1  & 9.9 & 9.1 & 8.7\\ 
0534292-051439 & 28.2 $\pm$ 1.3  & 25.3 $\pm$ 1.3  & 26.3 $\pm$ 1.5  & 15.2 & 15.4 & 13.6\\ 
0534302-051148 & 0.0 $\pm$ 0.0  & 0.0 $\pm$ 0.0  & 27.1 $\pm$ 3.1  & 0.0 & 0.0 & 5.5\\ 
0534336-051436 & 25.4 $\pm$ 1.2  & 24.8 $\pm$ 1.0  & 26.5 $\pm$ 1.1  & 15.9 & 22.8 & 18.9\\ 
\enddata
\end{deluxetable}

%% file: tab10.tex
\begin{deluxetable}{lll}
\rotate
\tablewidth{0pt}

\tabletypesize{\scriptsize}
\tablecaption{MIKE Field D Velocity Data}
\tablehead{
  \colhead{2massID} &  \colhead{RV D\_1}  & \colhead{R D\_1}   \\
    \colhead{}    &  \colhead{(km s$^{-1}$)} &  \\

   }
\startdata
0534435-045136 & 36.9 $\pm$ 3.9  & 4.2  \\ 
0534441-044751 & 10.6 $\pm$ 3.3  & 5.1  \\ 
0534452-044758 & 33.2 $\pm$ 1.9  & 10.3  \\ 
0534494-044539 & 37.9 $\pm$ 1.0  & 25.4  \\ 
0534512-044757 & 35.7 $\pm$ 2.0  & 9.5  \\ 
\enddata
\end{deluxetable}

%% file: tab11.tex
\begin{deluxetable}{llllllllllllll}
\rotate
\tablewidth{0pt}

\tabletypesize{\scriptsize}
\tablecaption{Spectroscopic Binaries from Velocity Shifts}
\tablehead{
  \colhead{RA} & \colhead{Dec.} & \colhead{2massID} &  \colhead{$\overline{RV}$\tablenotemark{a}} &  \colhead{Max. $\Delta$v} & \colhead{log(P)} & \colhead{$\chi^2_r$} & \colhead{N\_obs\tablenotemark{b}} & \colhead{V} & \colhead{V - I} &  \colhead{K} & \colhead{K - [3.6]} & \colhead{SB\tablenotemark{c}} & \colhead{Field ID\tablenotemark{d}}\\
  \colhead{(J2000)} & \colhead{(J2000)} &  &  \colhead{(km s$^{-1}$)} &  \colhead{(km s$^{-1}$)} &  &  &  &  &  & \\
}
\startdata
05:33:26.396 & -05:16:40.62 & 0533263-051640 & 67.0 $\pm$ 12.7 & 17.6 $\pm$ 12.8 & -44.43 & 102.31 & 3 & 17.95 & 2.64  & 12.09 & 0.15 & 2 & F2-E1, F2-E2, F22\\ 
05:33:29.832 & -05:27:35.38 & 0533298-052735 & -19.7 $\pm$ 2.7 & 2.4 $\pm$ 2.8 & -4.25 & 16.24 & 2 & 0 & 0  & 12.43 & 0.14 & 1 & F3-E1, F21\\ 
05:33:45.47 & -05:36:32.40 & 0533454-053632 & 18.0 $\pm$ 4.8 & 3.8 $\pm$ 4.8 & -9.20 & 38.23 & 2 & 18.04 & 3.7  & 11.00 & 0.82 & 1 & S1\\ 
05:33:47.802 & -04:55:03.63 & 0533478-045503 & 25.1 $\pm$ 2.5 & 2.8 $\pm$ 2.6 & -7.13 & 16.41 & 3 & 0 & 0  & 11.91 & 0.16 & 1 & F1-E1, F1-E2, F11\\ 
05:33:49.54 & -05:36:20.52 & 0533495-053620 & 24.7 $\pm$ 5.8 & 6.4 $\pm$ 6.0 & -5.80 & 13.35 & 3 & 16.2 & 2.08  & 10.71 & 0.65 & 1 & F2-E1, F2-E2, S3\\ 
05:33:54.573 & -05:14:15.54 & 0533545-051415 & 49.1 $\pm$ 13.9 & 13.4 $\pm$ 14.0 & -64.59 & 291.31 & 2 & 16.93 & 2.81  & 10.64 & 0.22 & 1 & F3-E1, F21\\ 
05:34:00.417 & -04:36:15.03 & 0534004-043615 & 32.4 $\pm$ 8.4 & 7.3 $\pm$ 9.3 & -10.63 & 24.48 & 3 & 0 & 0  & 11.80 & 0.26 & 2 & F1-E1, F1-E2, F11\\ 
05:34:07.110 & -05:49:25.19 & 0534071-054925 & 39.7 $\pm$ 8.1 & 8.0 $\pm$ 8.3 & -6.12 & 24.46 & 2 & 17.25 & 2.35  & 12.05 & 0.15 & 1 & F3-E1, F31\\ 
05:34:10.45 & -04:50:35.16 & 0534104-045035 & 23.8 $\pm$ 2.3 & 2.8 $\pm$ 2.5 & -7.41 & 12.45 & 4 & 13.01 & 1.2  & 9.95 & 0.9 & 1 & F1-E1, F1-E2, S2\\ 
05:34:12.029 & -05:24:19.63 & 0534120-052419 & 22.7 $\pm$ 5.4 & 3.8 $\pm$ 5.5 & -5.13 & 20.10 & 2 & 18.81 & 3.5  & 11.57 & 0.35 & 1 & B-1, B-2, F2-E1, F2-E2, F22\\ 
05:34:19.674 & -05:02:29.47 & 0534196-050229 & 40.4 $\pm$ 5.3 & 8.6 $\pm$ 5.3 & -18.34 & 29.50 & 4 & 17.86 & 2.82  & 11.60 & 0.22 & 2 & F2-E1, F2-E2, F3-E1, F22\\ 
05:34:20.315 & -04:34:03.43 & 0534203-043403 & 26.3 $\pm$ 2.6 & 2.6 $\pm$ 2.7 & -8.49 & 19.54 & 3 & 16.72 & 2.7  & 11.57 & 0.22 & 1 & F1-E1, F1-E2, F11\\ 
05:34:22.078 & -05:01:34.23 & 0534220-050134 & 22.7 $\pm$ 3.9 & 3.0 $\pm$ 4.0 & -5.82 & 23.15 & 2 & 18.45 & 3.16  & 12.25 & 0.54 & 1 & F3-E1, F22\\ 
05:34:24.246 & -04:24:39.99 & 0534242-042439 & 8.8 $\pm$ 3.6 & 5.0 $\pm$ 3.7 & -32.76 & 75.44 & 3 & 0 & 0  & 12.48 & 0 & 1 & F1-E1, F1-E2, F11\\ 
05:34:26.741 & -04:50:45.60 & 0534267-045045 & 32.1 $\pm$ 3.4 & 3.2 $\pm$ 3.4 & -15.36 & 35.36 & 3 & 17.04 & 2.63  & 11.33 & 0.23 & 1 & F1-E1, F1-E2, F11\\ 
05:34:28.677 & -04:34:39.58 & 0534286-043439 & 19.5 $\pm$ 4.6 & 5.5 $\pm$ 5.8 & -4.59 & 10.57 & 3 & 16.4 & 2.27  & 11.85 & 0.14 & 1 & F1-E1, F1-E2, F11\\ 
05:34:29.24 & -05:14:39.84 & 0534292-051439 & 27.8 $\pm$ 2.0 & 4.2 $\pm$ 2.6 & -4.02 & 4.66 & 7 & 15.07 & 2.27  & 9.43 & 0.65 & 1 & C-1, C-2, C-3, F1-E1, F1-E2, S1\\ 
05:34:29.50 & -05:13:55.20 & 0534295-051355 & 26.0 $\pm$ 5.2 & 5.6 $\pm$ 5.4 & -8.06 & 18.57 & 3 & 18.58 & 2.92  & 10.43 & 1.01 & 1 & C-1, C-2, C-3, F2-E1, F2-E2, S2\\ 
05:34:32.037 & -05:11:24.84 & 0534320-051124 & 48.3 $\pm$ 27.0 & 31.1 $\pm$ 27.0 & -163.89 & 377.37 & 3 & 19.95 & 4.55  & 11.84 & 0.37 & 1 & C-1, C-2, C-3, F2-E1, F2-E2, F21\\ 
05:34:33.012 & -05:57:47.06 & 0534330-055747 & 47.0 $\pm$ 19.8 & 15.1 $\pm$ 19.9 & -21.97 & 96.13 & 2 & 16.38 & 2.37  & 11.19 & 0.17 & 1 & A-1, A-2, F31\\ 
05:34:37.531 & -05:41:17.35 & 0534375-054117 & 29.2 $\pm$ 4.2 & 5.3 $\pm$ 4.4 & -6.75 & 15.53 & 3 & 15.92 & 2.58  & 10.31 & 0.24 & 1 & F2-E1, F2-E2, F31\\ 
05:34:39.039 & -04:55:28.83 & 0534390-045528 & 40.4 $\pm$ 12.3 & 15.7 $\pm$ 12.4 & -145.94 & 226.05 & 4 & 17.12 & 3.02  & 10.87 & 0.17 & 1 & F2-E1, F2-E2, F11\\ 
05:34:44.447 & -05:56:14.87 & 0534444-055614 & 25.8 $\pm$ 22.7 & 22.7 $\pm$ 22.8 & -28.38 & 125.40 & 2 & 17.12 & 2.57  & 12.24 & 0.15 & 1 & A-1, A-2, F31\\ 
05:34:45.244 & -04:47:58.11 & 0534452-044758 & 36.3 $\pm$ 6.6 & 8.2 $\pm$ 6.6 & -41.06 & 94.53 & 3 & 17.03 & 2.55  & 11.07 & 0.24 & 1 & D-2, F1-E1, F1-E2, F11\\ 
05:34:47.53 & -05:57:56.88 & 0534475-055756 & 9.9 $\pm$ 31.7 & 37.4 $\pm$ 31.8 & -192.57 & 297.72 & 4 & 14.62 & 1.67  & 9.40 & 0.69 & 1 & A-1, A-2, S3\\ 
05:34:49.27 & -06:00:11.30 & 0534492-060011 & 71.0 $\pm$ 13.9 & 11.9 $\pm$ 14.1 & -9.93 & 41.50 & 2 & 16.35 & 2.49  & 11.43 & 0.29 & 1 & A-1, A-2\\ 
05:34:51.754 & -05:39:24.12 & 0534517-053924 & 23.7 $\pm$ 3.7 & 3.7 $\pm$ 3.8 & -4.28 & 16.38 & 2 & 16.42 & 2.8  & 10.78 & 0.38 & 1 & F3-E1, F21\\ 
05:34:52.21 & -04:40:11.64 & 0534522-044011 & 65.6 $\pm$ 2.7 & 2.6 $\pm$ 3.2 & -20.61 & 33.03 & 4 & 14.38 & 1.62  & 9.80 & 0.73 & 1 & F1-E1, F1-E2, S2\\ 
05:34:55.603 & -05:29:37.60 & 0534556-052937 & 25.5 $\pm$ 6.5 & 6.0 $\pm$ 6.6 & -10.01 & 41.85 & 2 & 18.19 & 3.26  & 12.05 & 0.31 & 1 & B-1, B-2, F2-E1, F2-E2, F21\\ 
05:34:56.136 & -05:06:01.76 & 0534561-050601 & 29.9 $\pm$ 9.6 & 13.3 $\pm$ 10.3 & -4.16 & 9.57 & 3 & 16.87 & 2.77  & 11.67 & 0.27 & 2 & C-1, C-2, C-3, F1-E1, F1-E2, F11\\ 
05:34:59.322 & -05:05:30.03 & 0534593-050530 & 28.8 $\pm$ 2.1 & 2.7 $\pm$ 2.2 & -5.64 & 12.99 & 3 & 16.53 & 2.45  & 11.26 & 0.29 & 1 & C-1, C-2, C-3, F2-E1, F2-E2, F21\\ 
05:35:02.096 & -05:15:37.46 & 0535020-051537 & 28.4 $\pm$ 1.9 & 2.4 $\pm$ 2.0 & -4.11 & 7.22 & 4 & 16.5 & 2.37  & 11.79 & 0.48 & 1 & B-1, B-2, F1-E1, F1-E2, F21\\ 
05:35:03.91 & -05:29:03.48 & 0535039-052903 & 25.6 $\pm$ 15.8 & 23.2 $\pm$ 15.9 & -240.59 & 371.51 & 4 & 14.98 & 2.25  & 10.12 & 0.64 & 2 & B-1, B-2, F2-E1, F2-E2, S3\\ 
05:35:04.63 & -05:09:55.70 & 0535046-050955 & 29.7 $\pm$ 17.5 & 23.6 $\pm$ 17.6 & -30.29 & 69.75 & 3 & 15.73 & 2.79  & 10.39 & 0.18 & 2 & C-1, C-2, C-3\\ 
05:35:05.040 & -04:32:33.45 & 0535050-043233 & 31.9 $\pm$ 6.5 & 6.8 $\pm$ 6.7 & -10.45 & 24.07 & 3 & 17.76 & 3.08  & 12.03 & 0.34 & 1 & F1-E1, F1-E2, F11\\ 
05:35:05.21 & -05:14:50.28 & 0535052-051450 & 20.7 $\pm$ 7.4 & 9.6 $\pm$ 8.8 & -6.96 & 11.75 & 4 & 11.06 & 1.63  & 7.19 & 0.2 & 1 & C-1, C-2, C-3, F1-E1, F1-E2, S1\\ 
05:35:05.609 & -05:18:24.85 & 0535056-051824 & 6.2 $\pm$ 21.4 & 21.2 $\pm$ 21.5 & -34.35 & 152.71 & 2 & 15.53 & 0  & 11.25 & 0.39 & 1 & B-1, B-2, C-1, C-2, C-3, F1-E1, F1-E2, F2-E1, F2-E2, F22\\ 
05:35:11.657 & -05:31:01.15 & 0535116-053101 & 22.1 $\pm$ 6.6 & 8.7 $\pm$ 6.7 & -14.33 & 33.00 & 3 & 18.09 & 3.25  & 11.85 & 0.61 & 1 & B-1, B-2, F3-E1, F21\\ 
05:35:12.146 & -05:31:38.85 & 0535121-053138 & 49.5 $\pm$ 17.9 & 16.5 $\pm$ 18.0 & -66.60 & 153.36 & 3 & 16.95 & 3.11  & 11.12 & 0.22 & 1 & B-1, B-2, F2-E1, F2-E2, F22\\ 
05:35:12.33 & -05:54:35.20 & 0535123-055435 & 25.6 $\pm$ 6.1 & 4.8 $\pm$ 6.2 & -7.05 & 28.59 & 2 & 0 & 0  & 13.03 & 0.35 & 1 & A-1, A-2\\ 
05:35:15.953 & -05:14:59.04 & 0535159-051459 & 27.3 $\pm$ 1.5 & 1.9 $\pm$ 1.6 & -4.71 & 10.84 & 3 & 14.57 & 1.78  & 10.32 & 0.42 & 1 & B-1, B-2, F2-E1, F2-E2, F21\\ 
05:35:16.30 & -05:32:02.40 & 0535163-053202 & 21.4 $\pm$ 8.4 & 11.2 $\pm$ 8.5 & -14.38 & 33.11 & 3 & 18.66 & 3.61  & 11.89 & 0.66 & 1 & B-1, B-2, F3-E1, S3\\ 
05:35:17.95 & -05:26:50.64 & 0535179-052650 & 35.9 $\pm$ 21.4 & 18.5 $\pm$ 21.5 & -39.25 & 175.14 & 2 & 18.72 & 2.83  & 11.72 & 0.83 & 1 & C-1, C-2, C-3, S3\\ 
05:35:20.071 & -05:45:52.68 & 0535200-054552 & 20.0 $\pm$ 22.8 & 31.2 $\pm$ 23.1 & -19.50 & 44.91 & 3 & 0 & 0  & 12.14 & 0.16 & 1 & A-1, A-2, F22\\ 
05:35:22.18 & -05:24:24.80 & 0535221-052424 & 39.4 $\pm$ 28.8 & 39.0 $\pm$ 28.9 & -85.99 & 197.99 & 3 & 13.91 & 1.89  & 9.22 & 0.41 & 2 & C-1, C-2, C-3\\ 
05:35:22.63 & -05:14:11.04 & 0535226-051411 & 35.4 $\pm$ 8.4 & 12.3 $\pm$ 8.7 & -11.95 & 19.56 & 4 & 16.83 & 3.1  & 10.29 & 0.77 & 2 & C-1, C-2, C-3, F3-E1, S3\\ 
05:35:24.61 & -05:11:29.76 & 0535246-051129 & 28.3 $\pm$ 5.0 & 6.4 $\pm$ 5.2 & -8.19 & 13.68 & 4 & 18.9 & 4  & 10.39 & 0.75 & 1 & C-1, C-2, C-3, F2-E1, F2-E2, S1\\ 
05:35:26.84 & -05:11:07.44 & 0535268-051107 & 30.8 $\pm$ 3.0 & 5.0 $\pm$ 4.2 & -9.86 & 12.99 & 5 & 12.26 & 1.92  & 7.02 & 1.22 & 1 & C-1, C-2, C-3, F1-E1, F1-E2, S2\\ 
05:35:27.44 & -05:26:28.10 & 0535274-052628 & 26.7 $\pm$ 4.0 & 3.9 $\pm$ 4.1 & -4.36 & 10.04 & 3 & 14.12 & 1.75  & 9.96 & 0.17 & 1 & C-1, C-2, C-3\\ 
05:35:28.167 & -05:00:49.57 & 0535281-050049 & 42.8 $\pm$ 14.1 & 13.4 $\pm$ 14.8 & -16.83 & 38.76 & 3 & 16.85 & 2.75  & 11.54 & 0.25 & 1 & D-2, F3-E1, F22\\ 
05:35:29.896 & -05:12:10.39 & 0535298-051210 & 60.9 $\pm$ 15.2 & 25.6 $\pm$ 15.2 & -113.85 & 176.71 & 4 & 17.29 & 2.98  & 11.74 & 0.47 & 2 & C-1, C-2, C-3, F1-E1, F1-E2, F3-E1, F21\\ 
05:35:30.44 & -06:05:00.70 & 0535304-060500 & 38.8 $\pm$ 6.1 & 4.7 $\pm$ 6.3 & -6.21 & 24.85 & 2 & 19.68 & 3.21  & 13.53 & 0.6 & 2 & A-1, A-2\\ 
05:35:30.930 & -05:55:42.34 & 0535309-055542 & 24.1 $\pm$ 6.0 & 7.9 $\pm$ 6.2 & -14.34 & 23.29 & 4 & 0 & 0  & 12.40 & 0.51 & 1 & A-1, A-2, F31\\ 
05:35:39.923 & -04:58:39.27 & 0535399-045839 & -26.5 $\pm$ 57.5 & 57.4 $\pm$ 57.6 & -309.99 & 1419.84 & 2 & 16.8 & 2.64  & 11.78 & 0.2 & 1 & D-2, F11\\ 
05:35:42.76 & -05:11:54.70 & 0535427-051154 & 18.5 $\pm$ 9.8 & 10.2 $\pm$ 10.1 & -7.91 & 18.22 & 3 & 14.01 & 1.71  & 10.29 & 0.11 & 1 & C-1, C-2, C-3\\ 
05:35:43.061 & -05:03:07.55 & 0535430-050307 & 24.9 $\pm$ 4.6 & 5.7 $\pm$ 4.9 & -5.79 & 9.88 & 4 & 18.15 & 3.06  & 12.28 & 0.4 & 1 & F1-E1, F1-E2, F11\\ 
05:35:45.096 & -04:51:41.84 & 0535450-045141 & 20.6 $\pm$ 12.0 & 10.0 $\pm$ 12.1 & -14.12 & 60.45 & 2 & 0 & 0  & 12.04 & 0.28 & 1 & D-2, F1-E1, F1-E2, F11\\ 
05:35:45.318 & -04:43:39.58 & 0535453-044339 & 27.7 $\pm$ 2.8 & 2.9 $\pm$ 3.3 & -12.36 & 20.20 & 4 & 16.58 & 2.68  & 10.89 & 0.18 & 2 & D-2, F1-E1, F1-E2, F11\\ 
05:35:50.46 & -05:28:35.04 & 0535504-052835 & 51.5 $\pm$ 8.2 & 6.3 $\pm$ 8.3 & -9.25 & 38.46 & 2 & 10.3 & 0.7  & 6.22 & 0.6 & 2 & F2-E1, F2-E2, S3\\ 
05:35:51.11 & -05:07:08.76 & 0535511-050708 & 25.7 $\pm$ 1.5 & 1.9 $\pm$ 1.6 & -4.68 & 8.13 & 4 & 14.56 & 2.14  & 9.87 & 0.38 & 1 & C-1, C-2, C-3, F1-E1, F1-E2, S2\\ 
05:35:57.519 & -05:39:51.39 & 0535575-053951 & 63.5 $\pm$ 25.7 & 35.1 $\pm$ 25.9 & -30.79 & 70.89 & 3 & 0 & 0  & 11.71 & 0.19 & 2 & F2-E1, F2-E2, F21\\ 
05:35:58.973 & -05:59:08.47 & 0535589-055908 & 25.9 $\pm$ 4.1 & 5.1 $\pm$ 4.3 & -4.40 & 10.13 & 3 & 16.63 & 2.34  & 12.09 & 0.21 & 1 & A-1, A-2, F31\\ 
05:36:16.96 & -05:11:42.72 & 0536169-051142 & 24.3 $\pm$ 2.3 & 2.8 $\pm$ 2.4 & -7.34 & 12.34 & 4 & 0 & 0  & 9.28 & 0.38 & 1 & F2-E1, F2-E2, S1\\ 
05:36:18.788 & -05:12:18.25 & 0536187-051218 & 25.6 $\pm$ 2.7 & 3.0 $\pm$ 2.9 & -4.80 & 8.31 & 4 & 0 & 0  & 12.05 & 0.2 & 1 & F2-E1, F2-E2, F11\\ 
05:36:26.909 & -05:25:05.59 & 0536269-052505 & 32.8 $\pm$ 11.8 & 11.2 $\pm$ 11.9 & -60.73 & 273.60 & 2 & 17.01 & 3.1  & 10.32 & 0.18 & 1 & F3-E1, F22\\ 
05:36:36.335 & -05:36:34.11 & 0536363-053634 & 27.2 $\pm$ 7.1 & 9.3 $\pm$ 7.2 & -11.55 & 26.59 & 3 & 0 & 0  & 11.29 & 0.14 & 2 & F2-E1, F2-E2, F21\\ 
05:37:00.169 & -05:14:11.70 & 0537001-051411 & 35.9 $\pm$ 4.5 & 5.5 $\pm$ 4.6 & -13.27 & 30.56 & 3 & 0 & 0  & 12.11 & 0.18 & 1 & F2-E1, F2-E2, F21\\ 
05:37:09.044 & -05:17:12.09 & 0537090-051712 & 47.5 $\pm$ 15.8 & 18.1 $\pm$ 16.5 & -16.64 & 38.32 & 3 & 0 & 0  & 10.72 & 0.18 & 2 & F2-E1, F2-E2, F21\\ 
\enddata
\tablenotetext{a}{Velocities from \citet{furesz2008} are included in average. For a measurement to be included in the average R must be $>$ 4.5.}
\tablenotetext{b}{Number of observations with R $>$ 4.5.}
\tablenotetext{c}{1: Single-lined binary, 2: Double-lined binary.}
\tablenotetext{d}{F11, F21, F22, F31, S3, S2, S1 correspond to observations presented in \citet{furesz2008}.}

\end{deluxetable}

%% file: tab12.tex
\begin{deluxetable}{llllllllllllll}
\rotate
\tablewidth{0pt}

\tabletypesize{\scriptsize}
\tablecaption{Spectroscopic Binaries From Cross Correlation}
\tablehead{
  \colhead{RA} & \colhead{Dec.} & \colhead{2massID} &  \colhead{$\overline{RV}$\tablenotemark{a}} &  \colhead{Est. $\Delta$v}  & \colhead{N\_obs\tablenotemark{b}} & \colhead{V} & \colhead{V - I} &  \colhead{K} & \colhead{K - [3.6]} & \colhead{SB\tablenotemark{c}} & \colhead{Confidence\tablenotemark{d}} & \colhead{Field ID\tablenotemark{e}}\\
  \colhead{(J2000)} & \colhead{(J2000)} &  &  \colhead{(km s$^{-1}$)} &  \colhead{(km s$^{-1}$)} &  &  &  &  &  & \\
}
\startdata
05:33:29.386 & -05:07:49.10 & 0533293-050749 & 31.4 $\pm$ 3.7 & 0.0  & 3 & 0 & 0 & 12.09 & 0.15  & 2 & 2 & F2-E1, F2-E2, F22 &  \\ 
05:33:35.714 & -05:09:23.50 & 0533357-050923 & 30.8 $\pm$ 2.2 & 0.0  & 3 & 16.87 & 2.71 & 11.41 & 0.25  & 2 & 2 & F2-E1, F2-E2, F21 &  \\ 
05:33:36.441 & -04:49:49.02 & 0533364-044949 & 23.8 $\pm$ 3.3 & 0.0  & 3 & 0 & 0 & 12.12 & 0.31  & 2 & 2 & F1-E1, F1-E2, F11 &  \\ 
05:34:21.590 & -05:10:13.95 & 0534215-051013 & 24.9 $\pm$ 62.5 & 111.3  & 3 & 17.03 & 2.31 & 11.86 & 0.18  & 2 & 1 & F2-E1, F2-E2, F22 &  \\ 
05:34:27.836 & -05:43:31.52 & 0534278-054331 & 24.1 $\pm$ 2.8 & 0.0  & 4 & 16.89 & 2.7 & 11.74 & 0.18  & 2 & 2 & F2-E1, F2-E2, F3-E1, F31 &  \\ 
05:34:28.52 & -05:24:57.96 & 0534285-052457 & 24.0 $\pm$ 0.1 & 0.0  & 2 & 13.99 & 1.63 & 9.95 & 0.64  & 2 & 2 & B-1, B-2, S1 &  \\ 
05:34:30.207 & -04:58:30.42 & 0534302-045830 & 21.9 $\pm$ 2.0 & 75.8  & 4 & 16.32 & 2.45 & 11.49 & 0.18  & 2 & 1 & F1-E1, F1-E2, F3-E1, F11 &  \\ 
05:34:35.681 & -05:35:52.09 & 0534356-053552 & 25.8 $\pm$ 0.7 & 0.0  & 2 & 17.19 & 2.47 & 12.41 & 0.19  & 2 & 2 & B-1, B-2, F21 &  \\ 
05:34:40.87 & -05:22:42.24 & 0534408-052242 & -19.5 $\pm$ 1.1 & 88.8  & 1 & 13.4 & 1.75 & 8.60 & 1.37  & 2 & 1 & B-1, B-2, F2-E1, F2-E2, S2 &  \\ 
05:34:49.98 & -05:18:44.64 & 0534499-051844 & 57.3 $\pm$ 2.1 & 0.0  & 2 & 10.19 & 0.9 & 7.30 & 1.28  & 2 & 2 & F2-E1, F2-E2, S1 &  \\ 
05:34:52.339 & -05:30:08.01 & 0534523-053008 & 25.6 $\pm$ 2.5 & 0.0  & 1 & 17.57 & 3.01 & 12.13 & 0.24  & 2 & 2 & B-1, B-2, F21 &  \\ 
05:34:52.763 & -05:00:50.91 & 0534527-050050 & 24.4 $\pm$ 0.2 & 22.9  & 3 & 17.46 & 2.86 & 12.16 & 0.24  & 2 & 1 & F1-E1, F1-E2, F11 &  \\ 
05:34:53.12 & -05:47:42.50 & 0534531-054742 & 0.0 $\pm$ 0.0 & 0.0  & 0 & 18.34 & 2.73 & 13.16 & 0.21  & 2 & 2 & A-1, A-2 &  \\ 
05:34:55.63 & -06:01:03.60 & 0534556-060103 & 14.7 $\pm$ 4.0 & 0.0  & 2 & 15.88 & 2.14 & 11.57 & 0.14  & 2 & 2 & A-1, A-2 &  \\ 
05:34:57.452 & -05:30:42.02 & 0534574-053042 & 29.8 $\pm$ 2.4 & 0.0  & 1 & 16.54 & 2.4 & 11.76 & 0.2  & 2 & 2 & B-1, B-2, F21 &  \\ 
05:34:59.05 & -05:44:29.76 & 0534590-054429 & 23.4 $\pm$ 2.4 & 27.8  & 4 & 0 & 0 & 10.31 & 0.8  & 2 & 1 & F2-E1, F2-E2, S3 &  \\ 
05:35:02.497 & -05:33:09.95 & 0535024-053309 & 26.0 $\pm$ 1.3 & 0.0  & 2 & 15.67 & 2.3 & 11.00 & 0.84  & 2 & 2 & B-1, B-2, F3-E1, F22 &  \\ 
05:35:02.84 & -05:51:03.10 & 0535028-055103 & 21.6 $\pm$ 2.8 & 0.0  & 1 & 16.34 & 2.24 & 11.85 & 0.15  & 2 & 2 & A-1, A-2 &  \\ 
05:35:03.036 & -04:59:59.80 & 0535030-045959 & 26.1 $\pm$ 1.5 & 0.0  & 3 & 17.11 & 2.89 & 11.72 & 0.29  & 2 & 2 & D-2, F2-E1, F2-E2, F21 &  \\ 
05:35:04.76 & -05:17:42.10 & 0535047-051742 & 17.6 $\pm$ 2.5 & 0.0  & 1 & 17.35 & 3.13 & 9.34 & 1.1  & 2 & 2 & C-1, C-2, C-3 &  \\ 
05:35:05.60 & -05:25:19.20 & 0535056-052519 & 27.3 $\pm$ 1.8 & 0.0  & 5 & 11.52 & 1.34 & 7.43 & 0.74  & 2 & 2 & B-1, B-2, F2-E1, F2-E2, S1 &  \\ 
05:35:06.08 & -05:52:39.30 & 0535060-055239 & 12.9 $\pm$ 2.7 & 0.0  & 1 & 0 & 0 & 11.72 & 0.16  & 2 & 2 & A-1, A-2 &  \\ 
05:35:06.58 & -05:59:51.40 & 0535065-055951 & 31.2 $\pm$ 2.4 & 0.0  & 1 & 0 & 0 & 12.64 & 0.25  & 2 & 2 & A-1, A-2 &  \\ 
05:35:06.828 & -05:10:38.53 & 0535068-051038 & 19.9 $\pm$ 1.1 & 0.0  & 1 & 17.63 & 3.11 & 11.60 & 0.24  & 2 & 2 & C-1, C-2, C-3, F2-E1, F2-E2, F21 &  \\ 
05:35:07.24 & -04:50:25.4 & 0535072-045025 & 0.0 $\pm$ 0.0 & 0.0  & 0 & 16.8 & 2.88 & 11.34 & 0.23  & 2 & 2 & D-2 &  \\ 
05:35:09.96 & -05:57:11.90 & 0535099-055711 & 27.0 $\pm$ 5.9 & 0.0  & 2 & 0 & 0 & 9.61 & 1.25  & 2 & 2 & A-1, A-2 &  \\ 
05:35:11.97 & -05:20:33.10 & 0535119-052033 & 0.0 $\pm$ 0.0 & 0.0  & 0 & 16.85 & 3.37 & 8.85 & 0.66  & 2 & 2 & C-1, C-2, C-3 &  \\ 
05:35:12.59 & -05:23:44.16 & 0535125-052344 & 25.2 $\pm$ 1.3 & 28.0  & 4 & 12.64 & 1.62 & 8.65 & 0.44  & 2 & 1 & B-1, B-2, F2-E1, F2-E2, S2 &  \\ 
05:35:12.90 & -05:59:38.50 & 0535129-055938 & 0.0 $\pm$ 0.0 & 0.0  & 0 & 16.36 & 2.29 & 11.85 & 0.13  & 2 & 2 & A-1, A-2 &  \\ 
05:35:14.668 & -05:08:52.08 & 0535146-050852 & 26.2 $\pm$ 2.2 & 0.0  & 4 & 17.48 & 3 & 11.96 & 0.22  & 2 & 2 & C-1, C-2, C-3, F1-E1, F1-E2, F2-E1, F2-E2, F11 &  \\ 
05:35:15.758 & -05:21:39.82 & 0535157-052139 & 0.0 $\pm$ 0.0 & 0.0  & 0 & 17.68 & 2.99 & 11.38 & 0  & 2 & 2 & B-1, B-2, F21 &  \\ 
05:35:16.408 & -04:58:02.00 & 0535164-045802 & 24.2 $\pm$ 0.6 & 21.5  & 3 & 17.73 & 2.92 & 12.34 & 0.28  & 2 & 1 & D-2, F1-E1, F1-E2, F22 &  \\ 
05:35:17.139 & -05:12:39.41 & 0535171-051239 & 31.7 $\pm$ 6.6 & 0.0  & 3 & 18.74 & 3.44 & 12.60 & 0.3  & 2 & 2 & C-1, C-2, C-3, F21 &  \\ 
05:35:17.92 & -05:15:32.76 & 0535179-051532 & 28.9 $\pm$ 0.8 & 29.3  & 4 & 17.54 & 3.37 & 9.51 & 0.88  & 2 & 1 & F1-E1, F1-E2, S1 &  \\ 
05:35:18.098 & -04:31:19.33 & 0535180-043119 & 30.2 $\pm$ 2.6 & 0.0  & 3 & 17.14 & 2.83 & 11.79 & 0.21  & 2 & 2 & F1-E1, F1-E2, F11 &  \\ 
05:35:20.04 & -05:21:05.90 & 0535200-052105 & 20.3 $\pm$ 1.3 & 0.0  & 2 & 15.09 & 2.2 & 8.71 & 1.09  & 2 & 2 & C-1, C-2, C-3 &  \\ 
05:35:22.817 & -05:44:42.89 & 0535228-054442 & 42.7 $\pm$ 8.6 & 0.0  & 1 & 0 & 0 & 10.64 & 0.2  & 2 & 2 & F2-E1, F2-E2, F31 &  \\ 
05:35:23.23 & -04:43:02.9 & 0535232-044302 & 0.0 $\pm$ 0.0 & 25.7  & 0 & 15.08 & 2.06 & 10.77 & 0.09  & 2 & 1 & D-2 &  \\ 
05:35:23.97 & -05:59:41.90 & 0535239-055941 & 0.0 $\pm$ 0.0 & 0.0  & 0 & 13.77 & 1.56 & 10.33 & 0.11  & 2 & 2 & A-1, A-2 &  \\ 
05:35:24.25 & -05:25:18.48 & 0535242-052518 & 26.3 $\pm$ 1.6 & 0.0  & 1 & 14.85 & 1.98 & 10.04 & 0.43  & 2 & 2 & C-1, C-2, C-3, F3-E1, S1 &  \\ 
05:35:24.43 & -05:24:39.80 & 0535244-052439 & 21.6 $\pm$ 2.8 & 0.0  & 1 & 15 & 2.15 & 8.91 & 1.13  & 2 & 2 & C-1, C-2, C-3 &  \\ 
05:35:25.40 & -05:51:08.64 & 0535254-055108 & 26.9 $\pm$ 0.9 & 0.0  & 2 & 0 & 0 & 8.37 & 1.28  & 2 & 2 & A-1, A-2, S3 &  \\ 
05:35:25.676 & -04:57:18.35 & 0535256-045718 & 23.9 $\pm$ 4.6 & 40.0  & 4 & 17.92 & 3 & 12.37 & 0.25  & 2 & 1 & D-2, F1-E1, F1-E2, F11 &  \\ 
05:35:26.06 & -05:08:37.90 & 0535260-050837 & 28.2 $\pm$ 3.4 & 0.0  & 3 & 11.98 & 1.08 & 11.57 & 0  & 2 & 2 & C-1, C-2, C-3 &  \\ 
05:35:26.417 & -05:55:26.68 & 0535264-055526 & 24.9 $\pm$ 3.8 & 0.0  & 3 & 0 & 0 & 12.28 & 0.52  & 2 & 2 & A-1, A-2, F31 &  \\ 
05:35:26.589 & -04:56:06.71 & 0535265-045606 & 29.3 $\pm$ 1.3 & 40.2  & 4 & 15.39 & 2.14 & 10.93 & 0.25  & 2 & 1 & D-2, F1-E1, F1-E2, F2-E1, F2-E2, F22 &  \\ 
05:35:28.60 & -04:55:03.6 & 0535286-045503 & 75.6 $\pm$ 2.8 & 74.0  & 1 & 12.68 & 1.01 & 10.29 & -0.02  & 2 & 1 & D-2 &  \\ 
05:35:29.307 & -05:45:38.17 & 0535293-054538 & 26.0 $\pm$ 0.0 & 24.7  & 2 & 0 & 0 & 12.02 & 0.18  & 2 & 1 & A-1, A-2, F3-E1, F31 &  \\ 
05:35:30.48 & -05:24:23.00 & 0535304-052423 & 0.0 $\pm$ 0.0 & 0.0  & 0 & 14.37 & 1.89 & 10.08 & 0.23  & 2 & 2 & C-1, C-2, C-3 &  \\ 
05:35:30.913 & -05:18:17.95 & 0535309-051817 & 24.8 $\pm$ 1.1 & 0.0  & 3 & 16.39 & 3 & 10.41 & 0.36  & 2 & 2 & B-1, B-2, F2-E1, F2-E2, F21 &  \\ 
05:35:32.34 & -05:18:07.80 & 0535323-051807 & 32.0 $\pm$ 5.0 & 56.8  & 2 & 15.88 & 2.62 & 10.28 & 0.18  & 2 & 1 & C-1, C-2, C-3 &  \\ 
05:35:32.52 & -05:26:10.40 & 0535325-052610 & 0.0 $\pm$ 0.0 & 0.0  & 0 & 15.13 & 2.11 & 10.57 & 0.19  & 2 & 2 & C-1, C-2, C-3 &  \\ 
05:35:33.14 & -05:47:07.44 & 0535331-054707 & 23.9 $\pm$ 0.7 & 0.0  & 2 & 0 & 0 & 12.18 & 0.32  & 2 & 2 & A-1, A-2, S3 &  \\ 
05:35:33.456 & -04:56:01.76 & 0535334-045601 & 30.4 $\pm$ 1.9 & 0.0  & 2 & 17.91 & 3.1 & 12.06 & 0.22  & 2 & 2 & D-2, F1-E1, F1-E2, F22 &  \\ 
05:35:33.85 & -05:48:21.10 & 0535338-054821 & 15.3 $\pm$ 1.5 & 0.0  & 2 & 0 & 0 & 12.90 & 0.23  & 2 & 2 & A-1, A-2 &  \\ 
05:35:34.093 & -04:32:37.01 & 0535340-043237 & 32.4 $\pm$ 5.5 & 0.0  & 3 & 17.5 & 2.99 & 11.91 & 0.24  & 2 & 2 & F1-E1, F1-E2, F11 &  \\ 
05:35:37.341 & -06:00:00.21 & 0535373-060000 & 27.2 $\pm$ 0.4 & 0.0  & 1 & 17.26 & 2.63 & 12.21 & 0.24  & 2 & 2 & A-1, A-2, F31 &  \\ 
05:35:37.341 & -06:00:00.21 & 0535373-060000 & 27.2 $\pm$ 0.4 & 0.0  & 1 & 17.26 & 2.63 & 12.21 & 0.25  & 2 & 2 & A-1, A-2, F31 &  \\ 
05:35:38.214 & -05:14:19.02 & 0535382-051419 & 29.3 $\pm$ 1.2 & 0.0  & 3 & 16.1 & 2.86 & 10.52 & 0.16  & 2 & 2 & F2-E1, F2-E2, F21 &  \\ 
05:35:39.07 & -05:08:56.40 & 0535390-050856 & 14.2 $\pm$ 6.5 & 0.0  & 2 & 13.49 & 1.54 & 10.00 & 0.08  & 2 & 2 & C-1, C-2, C-3 &  \\ 
05:35:39.495 & -04:40:19.42 & 0535394-044019 & 31.1 $\pm$ 1.1 & 0.0  & 3 & 16.8 & 2.46 & 11.28 & 0.14  & 2 & 2 & D-2, F1-E1, F1-E2, F11 &  \\ 
05:35:42.84 & -05:51:36.80 & 0535428-055136 & 24.8 $\pm$ 2.4 & 0.0  & 2 & 0 & 0 & 11.74 & 0.23  & 2 & 2 & A-1, A-2 &  \\ 
05:35:43.24 & -05:09:17.10 & 0535432-050917 & 15.4 $\pm$ 7.6 & 0.0  & 3 & 13.79 & 1.74 & 9.85 & 0.09  & 2 & 2 & C-1, C-2, C-3 &  \\ 
05:35:48.853 & -05:00:28.56 & 0535488-050028 & 25.8 $\pm$ 5.7 & 0.0  & 5 & 17.22 & 2.57 & 11.28 & 0.14  & 2 & 2 & D-2, F1-E1, F1-E2, F2-E1, F2-E2, F11 &  \\ 
05:36:03.355 & -04:57:40.48 & 0536033-045740 & 23.9 $\pm$ 2.9 & 0.0  & 5 & 17.05 & 2.51 & 12.18 & 0.16  & 2 & 2 & D-2, F1-E1, F1-E2, F2-E1, F2-E2, F22 &  \\ 
05:36:05.655 & -05:52:13.08 & 0536056-055213 & 27.1 $\pm$ 7.2 & 23.2  & 2 & 0 & 0 & 11.97 & 0.22  & 2 & 1 & A-1, A-2, F31 &  \\ 
05:36:06.601 & -05:41:54.38 & 0536066-054154 & 27.2 $\pm$ 3.3 & 0.0  & 2 & 0 & 0 & 12.61 & 0.3  & 2 & 2 & F3-E1, F31 &  \\ 
\enddata
\tablenotetext{a}{Velocities from \citet{furesz2008} are included in average. For a measurement to be included in the average R must be $>$ 6.0.}
\tablenotetext{b}{Number of observations with R $>$ 6.0.}
\tablenotetext{c}{1: Single-lined binary, 2: Double-lined binary.}
\tablenotetext{d}{1: High confidence in detection, 2: moderate confidence in detection.}
\tablenotetext{e}{F11, F21, F22, F31, S3, S2, S1 correspond to observations presented in \citet{furesz2008}.}

\end{deluxetable}

%% file: tab13.tex
\begin{deluxetable}{llllll}

\tablewidth{0pt}

\tabletypesize{\scriptsize}
\tablecaption{Stellar and Gas Full Width Half-Maxima}
\tablehead{
  \colhead{Dec. of Bin} & \colhead{N} & \colhead{FWHM Stars} &\colhead{Median R} & \colhead{FWHM Stars} &\colhead{FWHM Gas}\\
  \colhead{}       & \colhead{Stars} & \colhead{Uncorrected} &                  & \colhead{Corrected} & \\
  \colhead{(J2000)}       &  & \colhead{(km s$^{-1}$)} &                  & \colhead{(km s$^{-1}$)} & \colhead{(km s$^{-1}$)}\\

   }
\startdata
-4.5$^{\circ}$ & 119 & 4.97 & 14.5 & 4.52 & 1.48 \\
-4.9$^{\circ}$ & 120 & 4.90 & 15.0 & 4.45 & 2.23 \\
-5.1$^{\circ}$ & 104 & 7.83 & 13.2 & 7.52 & 1.99 \\
-5.3$^{\circ}$ & 191 & 5.54 & 11.3 & 4.91 & 3.20 \\
-5.5$^{\circ}$ & 146 & 6.11 & 13.1 & 5.66 & 5.19 \\
-5.7$^{\circ}$ & 92  & 6.64 & 13.4 & 6.23 & 3.11 \\
-5.9$^{\circ}$ & 124 & 3.43 & 11.1 & 2.10 & 3.04 \\
-6.5$^{\circ}$ & 157 & 3.85 & 15.4 & 3.24 & 2.97 \\

\enddata

\end{deluxetable}

%% file: ms.bbl
\begin{thebibliography}{63}
\expandafter\ifx\csname natexlab\endcsname\relax\def\natexlab#1{#1}\fi

\bibitem[Allen et al.(2007)]{allen2007} Allen, L., et al.\ 2007, 
Protostars and Planets V, 361

\bibitem[Bernstein et al.(2003)]{bernstein2003} Bernstein, R., 
Shectman, S.~A., Gunnels, S.~M., Mochnacki, S., 
\& Athey, A.~E.\ 2003, \procspie, 4841, 1694 

\bibitem[Bally et al.(1987)]{bally1987} Bally, J., Stark, A.~A., 
Wilson, R.~W., \& Langer, W.~D.\ 1987, \apjl, 312, L45 

\bibitem[Bally et al.(1991)]{bally1991} Bally, J., Langer, W.~D., 
Wilson, R.~W., Stark, A.~A., 
\& Pound, M.~W.\ 1991, Fragmentation of Molecular Clouds and Star Formation, 147, 11 

\bibitem[{{Ballesteros-Paredes} {et~al.}(1999){Ballesteros-Paredes},
  {Hartmann}, \& {V{\'a}zquez-Semadeni}}]{bhv1999}
{Ballesteros-Paredes}, J., {Hartmann}, L., \& {V{\'a}zquez-Semadeni}, E. 1999,
  \apj, 527, 285

\bibitem[Batten(1973)]{batten1973} Batten, A.~H.\ 1973, Binary and multiple systems of stars


\bibitem[Bonnell et al.(2003)]{bonnell2003} Bonnell, I.~A., Bate, 
M.~R., \& Vine, S.~G.\ 2003, \mnras, 343, 413

\bibitem[Carpenter(2000)]{carpenter2000} Carpenter, J.~M.\ 2000, \aj, 
120, 3139 

\bibitem[Coelho et al.(2005)]{coelho2005} Coelho, P., Barbuy, B., Mel{\'e}ndez, J., Schiavon, R.~P., \& Castilho, B.~V.\ 2005, \aap, 443, 735 

\bibitem[Duquennoy \& Mayor(1991)]{dqmay1991} Duquennoy, A., \& Mayor, M.\ 1991, \aap, 248, 485 

\bibitem[Duch{\^e}ne(1999)]{duchene1999} Duch{\^e}ne, G.\ 1999, \aap, 341, 547 

\bibitem[Feigelson et al.(2005)]{fiegelson2005} Feigelson, E.~D., et al.\ 2005, \apjs, 160, 379

\bibitem[F{\H u}r{\'e}sz et al.(2006)]{furesz2006} F{\H 
u}r{\'e}sz, G., et al.\ 2006, \apj, 648, 1090

\bibitem[F{\H u}r{\'e}sz et al.(2008)]{furesz2008} F{\H 
u}r{\'e}sz, G., Hartmann, L.~W., Megeath, S.~T., Szentgyorgyi, A.~H., 
\& Hamden, E.~T.\ 2008, \apj, 676, 1109

\bibitem[Furlan et al.(2006)]{furlan2006} Furlan, E., et al.\ 
2006, \apjs, 165, 568 


\bibitem[Griffin et al.(1988)]{griffin1988} Griffin, R.~F., 
Griffin, R.~E.~M., Gunn, J.~E., \& Zimmerman, B.~A.\ 1988, \aj, 96, 172 

\bibitem[Hartmann et al.(1986)]{hartmann1986} Hartmann, L., Hewett, 
R., Stahler, S., \& Mathieu, R.~D.\ 1986, \apj, 309, 275 

\bibitem[{{Hartmann} {et~al.}(2001){Hartmann}, {Ballesteros-Paredes}, \&
  {Bergin}}]{hbb2001}
{Hartmann}, L., {Ballesteros-Paredes}, J., \& {Bergin}, E.~A. 2001, \apj, 562,
  852

\bibitem[Hartmann 
\& Burkert(2007)]{hb2007} Hartmann, L., \& Burkert, A.\ 2007, \apj, 654, 988 

\bibitem[Hillenbrand(1997)]{hill1997} Hillenbrand, L.~A.\ 1997, 
\aj, 113, 1733 

\bibitem[Hillenbrand 
\& Hartmann(1998)]{hh1998} Hillenbrand, L.~A., \& Hartmann, L.~W.\ 1998, \apj, 492, 540 

\bibitem[Huff 
\& Stahler(2007)]{huff2007} Huff, E.~M., \& Stahler, S.~W.\ 2007, \apj, 666, 281 

\bibitem[Ireland 
\& Kraus(2008)]{ireland2008} Ireland, M.~J., \& Kraus, A.~L.\ 2008, \apjl, 678, L59 

\bibitem[Jeffries(2008)]{jeffries2008} Jeffries, R.~D.\ 2008, 
arXiv:0811.3287 


\bibitem[Jones 
\& Walker(1988)]{joneswalker1988} Jones, B.~F., \& Walker, M.~F.\ 1988, \aj, 95, 1755 

\bibitem[Kerr 
\& Lynden-Bell(1986)]{kerr1986} Kerr, F.~J., \& Lynden-Bell, D.\ 1986, \mnras, 221, 1023 

\bibitem[Kraemer et al.(2003)]{kraemer2003} Kraemer, K.~E., 
Shipman, R.~F., Price, S.~D., Mizuno, D.~R., Kuchar, T., 
\& Carey, S.~J.\ 2003, \aj, 126, 1423 



\bibitem[Kroupa(2000)]{kroupa2000} Kroupa, P.\ 2000, New 
Astronomy, 4, 615 

\bibitem[Kutner et al.(1977)]{kutner1977} Kutner, M.~L., Tucker, 
K.~D., Chin, G., \& Thaddeus, P.\ 1977, \apj, 215, 521 


\bibitem[Lada 
\& Lada(2003)]{ladalada2003} Lada, C.~J., \& Lada, E.~A.\ 2003, \araa, 41, 57 

\bibitem[Larson(1981)]{larson1981} Larson, R.~B.\ 1981, \mnras, 
194, 809 



\bibitem[Mathieu et al.(1986)]{mathieu1986} Mathieu, R.~D., Latham, 
D.~W., Griffin, R.~F., \& Gunn, J.~E.\ 1986, \aj, 92, 1100 

\bibitem[Mathieu(1994)]{mathieu1994} Mathieu, R.~D.\ 1994, \araa, 32, 465 


\bibitem[Maxted et al.(2008)]{maxted2008} Maxted, P.~F.~L., 
Jeffries, R.~D., Oliveira, J.~M., Naylor, T., 
\& Jackson, R.~J.\ 2008, \mnras, 385, 2210 

\bibitem[McCabe et al.(2002)]{mccabe2002} McCabe, C., Duch{\^e}ne, 
G., \& Ghez, A.~M.\ 2002, \apj, 575, 974 

\bibitem[Mink 
\& Kurtz(1998)]{mink1998} Mink, D.~J., \& Kurtz, M.~J.\ 1998, Astronomical Data Analysis Software and Systems VII, 145, 93 

\bibitem[Munari et 
al.(2005)]{munari2005} Munari, U., Sordo, R., Castelli, F., \& Zwitter, T.\ 2005, \aap, 442, 1127 

\bibitem[Peterson 
\& Megeath(2008)]{peterson2008} Peterson, D.~E., \& Megeath, T.\ 2008, arXiv:0809.4006 

\bibitem[Prosser et al.(1994)]{prosser1994} Prosser, C.~F., 
Stauffer, J.~R., Hartmann, L., Soderblom, D.~R., Jones, B.~F., Werner, 
M.~W., \& McCaughrean, M.~J.\ 1994, \apj, 421, 517 


\bibitem[Proszkow et al. (2009)]{proszkow2008} Proszkow, E., 
Adams, F.~C., Hartmann, L., \& Tobin, J.~J.\ 2009, \apj ~in press. 

\bibitem[Rebull(2001)]{rebull2001} Rebull, L.~M.\ 2001, \aj, 121, 
1676 

\bibitem[Reipurth et al.(2007)]{reipurth2007} Reipurth, B., 
Guimar{\~a}es, M.~M., Connelley, M.~S., \& Bally, J.\ 2007, \aj, 134, 2272 

\bibitem[Roll(1996)]{roll1996} Roll, J.\ 1996, Astronomical Data 
Analysis Software and Systems V, 101, 536 

\bibitem[Scally et al.(2005)]{scm2005} Scally, A., Clarke, C., 
\& McCaughrean, M.~J.\ 2005, \mnras, 358, 742

\bibitem[Shu et 
al.(1987)]{shu1987} Shu, F.~H., Adams, F.~C., \& Lizano, S.\ 1987, \araa, 25, 23 

\bibitem[Sicilia-Aguilar et al.(2006)]{aurora2006} 
Sicilia-Aguilar, A., Hartmann, L.~W., F{\"u}r{\'e}sz, G., Henning, T., 
Dullemond, C., \& Brandner, W.\ 2006, \aj, 132, 2135 

\bibitem[Szentgyorgyi et al.(1998)]{szent1998} Szentgyorgyi, 
A.~H., Cheimets, P., Eng, R., Fabricant, D.~G., Geary, J.~C., Hartmann, L., 
Pieri, M.~R., \& Roll, J.~B.\ 1998, \procspie, 3355, 242 

\bibitem[Tan et al.(2006)]{tkm2006} Tan, J.~C., Krumholz, 
M.~R., \& McKee, C.~F.\ 2006, \apjl, 641, L121


\bibitem[Tonry \& Davis(1979)]{tonry1979} Tonry, J., \& Davis, M.\ 1979, \aj, 84, 1511 

\bibitem[van Altena et al.(1988)]{vanaltena1988} van Altena, W.~F., 
Lee, J.~T., Lee, J.-F., Lu, P.~K., \& Upgren, A.~R.\ 1988, \aj, 95, 1744 

\bibitem[Walker et al.(2007)]{walker2007} Walker, M.~G., Mateo, M., Olszewski, E.~W., Bernstein, R., Sen, B., \& Woodroofe, M.\ 2007, \apjs, 171, 389 



\end{thebibliography}
